\documentclass[twocolumn,journal]{IEEEtran}
\usepackage[T1]{fontenc}
\usepackage{array}
\usepackage{float}
\usepackage{textcomp}
\usepackage{amsmath}
\usepackage{amssymb}
\usepackage{graphicx}
\usepackage[unicode=true,
 bookmarks=true,bookmarksnumbered=true,bookmarksopen=true,bookmarksopenlevel=1,
 breaklinks=false,pdfborder={0 0 0},backref=false,colorlinks=false]
 {hyperref}
\hypersetup{pdftitle={Your Title},
 pdfauthor={Your Name},
 pdfpagelayout=OneColumn, pdfnewwindow=true, pdfstartview=XYZ, plainpages=false}
\usepackage{breakurl}

\makeatletter

\providecommand{\tabularnewline}{\\}

 \let\oldforeign@language\foreign@language
 \DeclareRobustCommand{\foreign@language}[1]{%
   \lowercase{\oldforeign@language{#1}}}

\usepackage[caption=false,font=footnotesize]{subfig}
\usepackage{array}
\usepackage[british,english]{babel}

\@ifundefined{showcaptionsetup}{}{%
 \PassOptionsToPackage{caption=false}{subfig}}
\usepackage{subfig}
\makeatother

\begin{document}

\title{Recovery of compressively sensed ultrasound images with structured
Sparse Bayesian Learning}

\author{Richard~Porter, Vladislav~Tadic, and~Alin~Achim\thanks{Richard~Porter is a PhD graduate of the Department of Electrical
\& Electronic Engineering, University of Bristol, Bristol, United
Kingdom, e-mail: \protect\href{mailto:eezrjp@my.bristol.ac.uk}{eezrjp@my.bristol.ac.uk}.}\thanks{Vladislav~Tadic is with the School of Mathematics, University of
Bristol, Bristol, United Kingdom, e-mail: \protect\href{http://v.b.tadic@bristol.ac.uk}{v.b.tadic@bristol.ac.uk}.}\thanks{Alin~Achim is with the Department of Electrical \& Electronic Engineering,
University of Bristol, Bristol, United Kingdom, e-mail: \protect\href{mailto:eeama@bristol.ac.uk}{eeama@bristol.ac.uk}.}}

\markboth{}{Richard Porter \MakeLowercase{\emph{et al.}}: Recovery of compressively
sensed ultrasound images with structured Sparse Bayesian Learning}
\maketitle
\begin{abstract}
In this paper, we consider the problem of recovering compressively
sensed ultrasound images. We build on prior work, and consider a number
of existing approaches that we consider to be the state-of-the-art.
The methods we consider take advantage of a number of assumptions
on the signals including those of temporal and spatial correlation,
block structure, prior knowledge of the support, and non-Gaussianity.
We conduct a series of intensive tests to quantify the performance
of these methods. We find that by altering the parameters of the structured
Sparse Bayesian Learning approaches considered, we can significantly
improve the objective quality of the reconstructed images. The results
we achieve are a significant improvement upon previously proposed
reconstruction techniques. In addition, we further show that by careful
choice of parameters, we can obtain near-optimal results whilst requiring
only a small fraction of the computational time needed for the best
reconstruction quality.\end{abstract}


\IEEEpeerreviewmaketitle{}

\section{Introduction}

Ultrasound imaging is possibly the most commonly used cross-sectional
medical imaging modality. It has a number of advantages over alternatives,
as it is relatively cheap, can easily be made portable, non-invasive
and does not make use of ionising radiation. Ultrasound can also produce
``real-time'' images, and it is generally considered to be safe
\cite{merritt1989ultrasound}.

In general, ultrasound images are formed by the transmission of short
ultrasound pulses from an array of transducers (most commonly piezoelectric
transducers) towards the object of interest \cite{szabo2004diagnostic}.
The returning (reflected) echoes are analysed and processed in order
to construct an image of the object being scanned. As with all imaging
modailities, ultrasound imaging generates a significant amount of
data. Therefore image compression is needed to reduce the volume of
data hence reducing the bit rate, and ideally this compression should
not lead to any loss in perceptual image quality. The need for storage
space and transmission bandwidth, particularly that caused by the
diversification of ultrasound applications and telemedicine, place
significant demands on existing systems in digital radiology departments
\cite{szabo2004diagnostic}. The development of new technologies,
allowing for the acquisition of ever greater amounts of data, places
even greater demands on data processing, transmission and storage
capabilities, giving rise to a need for more efficient compression
techniques. In the field of medical ultrasound, there have been several
recent developments that have significantly increased the amount of
data generated. These developments include scanners with the ability
to produce real-time 3D (RT3D) \cite{stetten1998real} or 4D \cite{yagel20073d}
image data sets. One issue with these techniques is that of low frame
rates, with most scanners capable of generating only a few images
per second. Whilst this is fast enough to view fetal facial expressions,
it is not fast enough to view the operation of the fetal heart in
detail. Although several techniques have been proposed to increase
the frame rates of these methods, such as multiline transmit imaging,
plane-/diverging wave imaging, and retrospective gating, acquiring
data at these higher frame rates results in a loss of image quality
\cite{cikes2014ultrafast,tanter2014ultrafast}. 

The phenomenon of growth in the amount of data being generated outstripping
the growth of data processing and storage capabilities is not unique
to the field of ultrasound or medical imaging in general \cite{baraniuk2011more}.
One approach that has been proposed to deal with this growth in data
is that of compressed sensing. The field of compressed sensing has
grown from work by Candès, Romberg, Tao \cite{candes2006robust},
and Donoho \cite{donoho2006compressed} on the single measurement
vector (SMV) model. Later work has shown that with the shared sparsity
assumption, performance can be increased in the multiple measurement
vector (MMV) model \cite{cotter2005sparse}.

Compressed sensing leverages the concept of sparsity, which is fundamental
to much of modern signal processing. The idea underlying this is that
many natural signals can be represented with less data than the number
of samples that would be implied by the Nyquist sampling theorem \cite{candes2006robust}.
This concept is used in transform coding, for example in JPEG \cite{wallace1991jpeg}
for image coding and MPEG \cite{le1991mpeg} for video coding. In
transform coding approaches, the signal must first be acquired at
the Nyquist rate, and then compressed, effectively wasting much of
the acquired data. The method of compressed sensing allows us to reduce
the rate at which we sample signals, thus avoiding the need to first
sample at the Nyquist rate by combining the acquisition step with
the compression step. This is achieved due to two significant differences
between compressed sensing and classical sampling \cite{baraniuk2011introduction}.
Firstly, rather than sampling at specific points in time as with classical
sensing, compressed sensing typically consists of taking inner products
between the signal and general sampling kernels. Secondly, signal
reconstruction in the Shannon-Nyquist framework is done by sinc interpolation,
and this takes very little computation, whereas compressed sensing
signal recovery methods are typically computationally intensive. In
addition, with traditional transform coding approaches, the quality
of the resulting image is determined primarily by the encoder at the
time of encoding, whereas with compressed sensing development of improved
recovery algorithms may improve the quality of the final image. The
ability of compressed sensing techniques to allow signals to be acquired
at rates below the Nyquist rate may allow for an increase in the framerate
of ultrasound imaging techniques by reducing the amount of data acquired.

In medical imaging. compressed sensing techniques have been successfully
applied to MRI in order to reduce scan time \cite{vasanawala2011practical}.
MRI is particularly amenable to compressed sensing techniques as the
images are already acquired in the Fourier domain (k-space), and therefore
the primary difficulty lies in the design of appropriate sampling
patterns, and does not require the development of any new hardware.
MRI is also of interest as it is the other commonly used cross-sectional
medical imaging modality that does not make use of ionising radiation.
Although MRI is capable of greater detail than ultrasound, it is significantly
more expensive, not portable, and has much slower scan times.

Typically, compressed sensing approaches make no assumption on the
signals being acquired other than that of sparsity. However, it is
often the case that we may have knowledge of the signal properties,
and the use of this knowledge can improve the ability to reconstruct
compressively sensed signals. For example, it may be the case that
we expect the signals to be temporally correlated, and a method was
proposed in \cite{zhang2011sparse} to reduce the negative effect
of temporal correlation on the recovery of compressively sensed signals.
In addition, we might expect the non-zero elements of a signal to
cluster together, and this leads to the assumption of block structure,
used in \cite{zhang2012recovery}, and combined with the assumption
of temporal correlation in \cite{zhang2014spatiotemporal}. Another
approach is to take into account the expected statistical properties
of the signals. Although a common assumption, justified by the central
limit theorem is that signals are Gaussian, in recent years it has
become known that some natural signals do not obey this assumption.
As the primary assumption needed for the central limit theorem to
be applicable is that of finite variance, it is not surprising that
we find that these signals can often be modeled as $\alpha-$stable
distributions, as is implied by the generalised central limit theorem
\cite{gnedenko1954limit}. These distributions have found applications
in financial modeling \cite{voit2013statistical} (indeed, it has
been argued that the 2007-2008 financial crisis can be partially attributed
to the model error caused by the assumption of Gaussianity \cite{marsh2012black}),
and it has also been shown that ultrasound images can be better modeled
by a symmetric $\alpha-$stable (S$\alpha$S) distribution than by
a Gaussian distribution \cite{kutay2001modeling}.

There has been significant work done on applying compressed sensing
techniques to ultrasound imaging. In 2012, \cite{liebgott2012compressive}
produced a review of these attempts which suggested that they fall
into four categories.

The first category consists of methods that model the object being
scanned as a sparse collection of scatterers. This is perhaps the
easiest form to implement with existing ultrasound hardware. \cite{schiffner2011fast}
demonstrated an implementation of this idea, although they did note
some difficulties with dealing with the sensing matrix (estimated
to be 458GB for a typical problem size), which they addressed by using
a powerful GPU and recomputing the entries of the sensing matrix instead
of storing them. It is also worth noting that they used a discrete
cosine basis, as is used in this paper (although they used a 2D DCT).
Work since includes the work of \cite{chen2015reconstruction}, who
modelled the acquired signals as a convolution of a point spread function
and tissue reflectivity function, and improved on the reconstruction
quality of previous work. The previous work of \cite{chen2016compressive}
combined deconvolution and CS ideas.

The second category consists of methods that take advantage of the
sparsity of the raw RF data, e.g. \cite{demanet2007wave} and \cite{liebgott2013pre}.
More recent work includes the work of \cite{wagner2012compressed},
who introduced the idea of compressed beamforming in the context of
the Xampling framework, and the work of \cite{chernyakova2014fourier},
which extended this work to the idea of beamforming in the frequency
domain.

The third category consists of methods that take advantage of the
sparsity of ultrasound images in the 2D Fourier transform domain.
Several of these such as \cite{quinsac2011bayesian} and \cite{dobigeon2012regularized}
adopt a Bayesian approach for the reconstruction of ultrasound images.\cite{achim2010compressive}
introduced a framework for the compressed sensing of medical ultrasound
based on modelling data with a S$\alpha$S distribution, and an approach
using the iteratively reweighted least squares (IRLS) algorithm for
$\ell_{p}$ pseudonorm minimisation was proposed, with $p$ related
to the the characteristic exponent of the distribution of the underlying
data. This approach was further extended in \cite{achim2014reconstruction},
where it was shown that performance could be improved by taking into
account knowledge of the support. Another approach using a line-by-line
strategy is in \cite{tzagkarakis2013joint}, which made use of the
correlations between each ultrasound line. It is these approaches
that are most closely related to the work presented in this paper.
Another 1D approach can be found in the work of \cite{tur2011innovation},
where an FRI based approach was used.

The final category relates to Doppler imaging, which is a problem
with a somewhat different nature \cite{lorintiu2016compressed,zobly2011compressed}.

We have previously shown that it is possible to improve the reconstruction
performance by taking advantage of the non-Gaussianity, temporal and
block structure of the ultrasound data \cite{PortICIP2015,porter2015sparse},
building on the work in \cite{tzagkarakis2013joint} which was the
first to apply the T-MSBL method (compensating for the negative effect
of temporal correlation) to the recovery of compressively sensed ultrasound
images. The acquisition of medical ultrasound data in a manner suitable
for compressed sensing techniques has been examined in other works,
e.g. \cite{wagner2012compressed}, and also general methods using
compressed sensing for sub Nyquist analog-to-digital convertors have
been developed, e.g. \cite{mishali2011xampling2}, but this is beyond
the scope of this paper. Here, we build on previous work, and show
that by careful selection of parameters for selected structured Sparse
Bayesian Learning methods, we can significantly improve the resulting
reconstruction quality. We also compare these  methods to a number
of existing approaches, and show that we obtain significant improvements. 

The rest of this paper is organised as follows. We first introduce
some technical background in section \ref{sec:background}, while
in section \ref{sec:methods} we introduce the methods we will be
comparing. In section \ref{sec:data} we describe the datasets used,
in section \ref{sec:results} we present our results on these datasets,
and in section \ref{sec:conclusions} we conclude the paper.

\section{Background}

\label{sec:background}

In this section, we provide a brief overview of the models and theory
used for compressed sensing, as well as introducing the notation used
in the paper.

\subsection{Notation}
\begin{itemize}
\item $\left\Vert \mathbf{x}\right\Vert _{0}$,$\left\Vert \mathbf{x}\right\Vert _{1}$,$\left\Vert \mathbf{x}\right\Vert _{2}$
denote the $\ell_{0}$ pseudo-norm, and the $\ell_{1}$ and $\ell_{2}$
norms of the vector $\mathbf{x}$. 
\item $\mathbf{A}_{i.}$ denotes the $i^{th}$ row of the matrix $\mathbf{A}$,
and $\mathbf{A}_{.j}$ denotes the $j^{th}$ column of the matrix
$\mathbf{A}$
\item $\mathbf{A}\otimes\mathbf{B}$ denotes the Kronecker product of matrices
$\mathbf{A}$ and \textbf{$\mathbf{B}$}
\end{itemize}

\subsection{Models}

The SMV model of compressed sensing is given by 

\begin{equation}
\mathbf{y}=\mathbf{A}\mathbf{x}+\mathbf{v}\label{eq:-7}
\end{equation}

Here, $\mathbf{y}\in\mathbb{R}^{N\times1}$ represents the observed
measurements, $\mathbf{A}\in\mathbb{R}^{N\times M}$ is the measurement
matrix, $\mathbf{v}\in\mathbb{R}^{N\times1}$ is a noise vector, and
$\mathbf{x}\in\mathbb{R}^{M}$ is the source vector we want to recover.
In the context of ultrasound imaging, we can consider $\mathbf{x}$
to correspond to a line of the ultrasound image, with each line corresponding
to a single transducer element. The elements of $\mathbf{x}$ therefore
correspond to equally spaced time domain samples of the reflected
echoes. 

The MMV model is given by 

\begin{equation}
\mathbf{Y}=\mathbf{A}\mathbf{X}+\mathbf{V}\label{eq:-13}
\end{equation}

Here, $\mathbf{Y}\in\mathbb{R}^{N\times L}$ represents the observed
measurements, $\mathbf{A}\in\mathbb{R}^{N\times M}$ is the measurement
(or sensing) matrix, $\mathbf{V}\in\mathbb{R}^{N\times L}$ is a noise
matrix, and $\mathbf{X}\in\mathbb{R}^{M\times L}$ is the source matrix
we want to recover, with each row corresponding to a possible source.
In the ultrasound context, $X$ is now the entire ultrasound image,
with each column of $\mathbf{X}$ corresponding to a single line of
the ultrasound image. The columns of $\mathbf{X}$ are arranged such
that they correspond to the spatial positions of the individual transducers.

Compressed sensing relies upon the idea of sparsity. We say that $\mathbf{x}$
is $k$-sparse if at most $k$ components of $\mathbf{x}$ are non-zero,
and similarly we will say that $\mathbf{X}$ is $k$-sparse if at
most $k$ rows of $\mathbf{X}$ are non-zero.

In the absence of noise, in the SMV case, it has been shown that under
certain assumptions (which are satisfied with probability 1 if the
entries of $\Phi$ are drawn independently from a continuous probability
distribution), $2k$ measurements (i.e. $N=2k$) are sufficient to
guarantee the exact recovery of $\mathbf{x}$ by finding the $x$
with the minimal number of non-zero elements such that $\Phi\mathbf{x}=\mathbf{y}$
\cite{baraniuk2011introduction}, and under similar assumptions, along
with the assumption that $X$ is of maximal column rank and has a
sufficient number of columns, $k+1$ measurements are sufficient to
recover $\mathbf{X}$ exactly by finding the $\mathbf{X}$ with the
minimal number of non-zero rows such that $\mathbf{AX}=\mathbf{Y}$
\cite{cotter2005sparse}.

However, this method of recovery is computationally expensive, as
it requires searching over the possible sets of non-zero elements
or columns. If we assume that there are at most $k$ non-zero elements
(or rows), and assume that we start searching from the smallest possible
set, then we would need to check $\mathcal{O}(M^{k})$ sets, as shown
in equation 

\begin{equation}
\sum_{j=0}^{k}\left(\begin{array}{c}
M\\
j
\end{array}\right)=\mathcal{\mathcal{O}}(M^{k})\label{eq:}
\end{equation}

It is clear that this quickly becomes unfeasible for larger values
of $M$ and $k$, and hence faster approximate methods such as $\ell_{1}$
norm minimisation are used. If we assume that the entries of the sensing
matrix are drawn independently from a Gaussian distribution with zero
mean \cite{candes2006stable} and variance $\frac{1}{N}$, then approximately
$kC\log(\frac{M}{N})$ measurements are needed to ensure that the
recovery of a $k$ sparse vector will be exact (i.e. the recovery
based on $\ell_{1}$ norm-minimisation will coincide with that of
the recovery based on $\ell_{0}$ pseudonorm minimisation) with high
probability, and similarly, that if a $k$-sparse vector provides
a good approximation of $\mathbf{x}$, $kC\log(\frac{M}{N})$ are
needed to ensure that the estimate of $\mathbf{x}$ recovered with
$\ell_{1}$ minimisation can also be expected to provide a good approximation
for $\mathbf{x}$.

\section{Reconstruction Methods}

\label{sec:methods}

This section describes several methods that can be used for the reconstruction
of compressively sensed signals. The methods include techniques for
both the SMV and MMV cases, and take into consideration assumptions
on the signals including those of temporal and spatial correlation,
block structure, prior knowledge of the support, and non-Gaussianity.

\subsection{Temporal-Multiple Sparse Bayesian Learning \& Temporal-Multiple Sparse
Bayesian Learning-Mixture of Gaussians-a}

The T-MSBL and T-MSBL-MoG-a algorithms for the MMV model are as described
in the work of \cite{zhang2011sparse} and \cite{porter2015sparse}.
The core idea is to learn the correlation structure between the measurement
vectors and compensate for it.

\subsection{Block Sparse Bayesian Learning-Bound Optimization}

Another technique derived in a similar way to T-MSBL is the method
of Block Sparse Bayesian Learning (BSBL) proposed by \cite{zhang2012recovery}.
This method can be used to exploit the fact that the non-zero components
of each sample in time (column of the image) tend to occur in clusters.
This technique works on each column individually as a technique for
the SMV model. 

The block structure model for $\mathbf{x}$ is given by equation (\ref{eq:-1-3}).

\begin{equation}
\mathbf{x}=[\underbrace{x_{1,}...x_{d_{1}}}_{\mathbf{x_{1}^{\text{T}}}},...,\underbrace{x_{d_{g-1}+1},...,x_{d_{g}}}_{\mathbf{x}_{g}^{\text{T}}}]^{\text{T}}\label{eq:-1-3}
\end{equation}

The assumption used is that each block is independent, and distributed
according to a zero-mean Gaussian distribution. This gives the prior
shown by equations (\ref{eq:-3-1-1}) and (\ref{eq:-12-3}).

\begin{equation}
p(\mathbf{x};\gamma_{i},B_{i},\forall i)\sim\mathcal{N}(\mathbf{0},\Sigma_{0})\label{eq:-3-1-1}
\end{equation}
\begin{equation}
\Sigma_{0}=\begin{bmatrix}\gamma_{1}\mathbf{B}_{1}\\
 & .\\
 &  & .\\
 &  &  & .\\
 &  &  &  & \gamma_{g}\mathbf{B}_{g}
\end{bmatrix}\label{eq:-12-3}
\end{equation}

Here, $\mathbf{B}_{i}$ represents the correlation structure within
a block, and $\gamma_{i}$ an unknown nonnegative scalar parameter
that determines the sparsity level of the $i$-th block. To avoid
overfitting, it is assumed that $\mathbf{B}_{i}=\mathbf{B}_{j}=\mathbf{B}\forall i,j$

In this paper, it will be assumed that the blocks all have equal length.

The learning rules are obtained by following an Expectation-Maximisation
method (for details, see the work of \cite{zhang2011sparse}). In
this paper, the BSBL-Bound Optimisation (BSBL-BO) algorithm, which
is significantly faster than the Expectation-Maximisation based BSBL
algorithm \cite{zhang2013extension} is used. This changes only the
learning rule for $\gamma_{i}$, the other learning rules remain the
same.

\subsection{Spatio Temporal-Sparse Bayesian Learning}

The ST-SBL method, proposed by \cite{zhang2014spatiotemporal} is
combination of the block sparsity idea of the BSBL method, and the
correction for temporal correlation of the T-MSBL method \cite{zhang2014spatiotemporal},
and it works on the MMV model.

The assumption ST-SBL makes on the structure of $\mathbf{X}$ is that
it has block structure as given by 

\begin{equation}
\mathbf{X}=\left[\begin{array}{c}
\mathbf{X}_{[1].}\\
\mathbf{X}_{[2].}\\
\vdots\\
\mathbf{X}_{[g].}
\end{array}\right]\label{eq:-15-2}
\end{equation}

Where $\mathbf{X}_{[i].}\in\mathbb{R}^{d_{i}\times L}$ is the $i$-th
block of $\mathbf{X}$ and $\sum_{i=1}^{g}d_{i}=N$, and it is assumed
that only a few of the blocks $\mathbf{X}_{[i].}$ are non-zero. As
with the BSBL-BO method, in this paper it will be assumed that the
blocks are all of equal size, and therefore $d_{i}=d\forall i$. For
each block, it is assumed that entries in the same row of $\mathbf{X}_{[i].}$
are correlated and that entries in the same column of $\mathbf{X}_{[i].}$
are correlated, and therefore that each block has spatiotemporal correlation.

Similarly to the other SBL based algorithms, it is assumed that each
block has a Gaussian distribution as

\begin{equation}
p(\text{vec}(\mathbf{X}_{[i].}^{T});\gamma_{i},\mathbf{B},\mathbf{C}_{i})=\mathcal{N}(0,(\gamma_{i}\mathbf{C}_{i})\otimes B)\label{eq:-35-1}
\end{equation}

Here, $\mathbf{B}\in\mathbb{R}^{L\times L}$ is an unknown positive
definite matrix that captures the correlation structure in each row
of $\mathbf{X}_{[i].}$, $\mathbf{C}_{i}\in\mathbb{R}^{d\times d}$
is an unknown positive definite matrix that captures the correlation
structure in each column of $\mathbf{X}_{[i].}$, and $\gamma_{i}$
is an unknown nonnegative scalar parameter that determines the sparsity
level of the $i$-th block.

The distribution of $\mathbf{X}$ (assuming independence of the blocks)
can be written as shown by equation (\ref{eq:-37-1}). 

\begin{equation}
p(\text{vec}(\mathbf{X}^{T});\mathbf{B},\{\mathbf{C}_{i},\gamma_{i}\}_{i})=\mathcal{N}(0,\Pi\otimes\mathbf{B})\label{eq:-37-1}
\end{equation}

Where $\Pi$ is a block diagonal matrix given by

\begin{equation}
\Pi\triangleq\left[\begin{array}{cccc}
\gamma_{1}\mathbf{C}_{1}\\
 & \gamma_{2}\mathbf{C}_{2}\\
 &  & \ddots\\
 &  &  & \gamma_{g}\mathbf{C}_{g}
\end{array}\right]\label{eq:-38-1}
\end{equation}

The relationship to the BSBL model is clear and indeed when $L=1$
the ST-SBL model reduces to the BSBL model.

As with BSBL and T-MSBL, the learning rules are found by following
an Expectation-Maximisation algorithm. For details, see the work of
\cite{zhang2014spatiotemporal}.

\medskip{}

Note that the implementations of both the ST-SBL and BSBL-BO algorithm
use a rule to remove blocks (i.e. set them to zero) if the corresponding
$\gamma_{i}$ is below a certain threshold. This is done by removing
the corresponding columns of $\mathbf{A}$ and rows of $\mathbf{X}$
to produce $\tilde{\mathbf{A}}$ and $\tilde{\mathbf{X}}$ which are
then used for the remainder of the process. In this paper, this threshold
will be referred to as $\bar{\Gamma}$.

\subsection{Iterative Reweighted Least Squares with Dual Prior}

It was shown by \cite{kutay2001modeling} that ultrasound RF echoes
can be modeled using a power-law shot noise model, and it was shown
by \cite{petropulu2000power} that this model is related to $\alpha$-stable
distributions. The IRLS approach for $\ell_{p}$ pseudonorm minimisation
has been used to attempt to take advantage of this, with $p$ related
to $\alpha$ \cite{achim2010compressive}. This has been further extended
to use knowledge of the support by \cite{achim2014reconstruction},
using the modified IRLS algorithm from the work of \cite{miosso2009compressive},
and it is that algorithm that will be used in this paper. $p$ is
related to the characteristic exponent of the underlying distribution
by $p=\alpha-0.01$.

\subsubsection{Block IRLS}

The BIRLS algorithm used in this paper is an adaptation of the IRLS
algorithm, with the weights calculated by summing across each block,
and no prior support information is used.

\section{Compressive ultrasound simulations}

\label{sec:data}

\subsection{Thyroid image data set}

\begin{figure}[H]
\begin{centering}
\subfloat[\label{fig:DCT-of-image1-1-1}]{\protect\includegraphics[width=1.6cm,height=6cm]{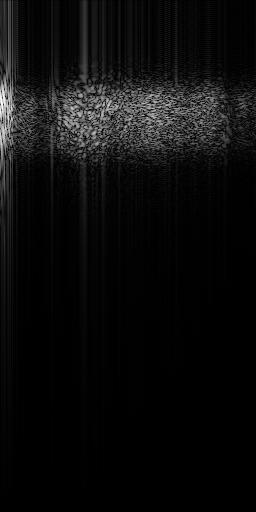}}\ \subfloat[\label{fig:DCT-of-image1-1-1}]{\protect\includegraphics[width=1.6cm,height=6cm]{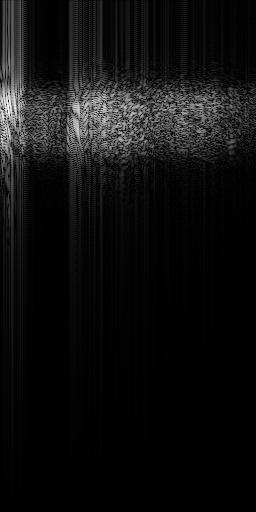}}\ \subfloat[\label{fig:DCT-of-image1-2-1}]{\protect\includegraphics[width=1.6cm,height=6cm]{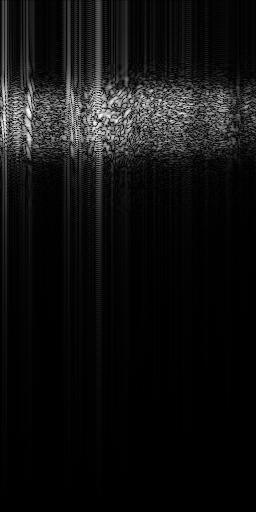}}
\par\end{centering}

\centering{}\protect\caption{\label{fig:DCT-of-image1-3}DCTs of images (a) 1, (b) 2 and (c) 3
from image set 1}
\end{figure}

\begin{figure}[H]
\begin{centering}
\subfloat[]{\protect\includegraphics[width=1.75cm]{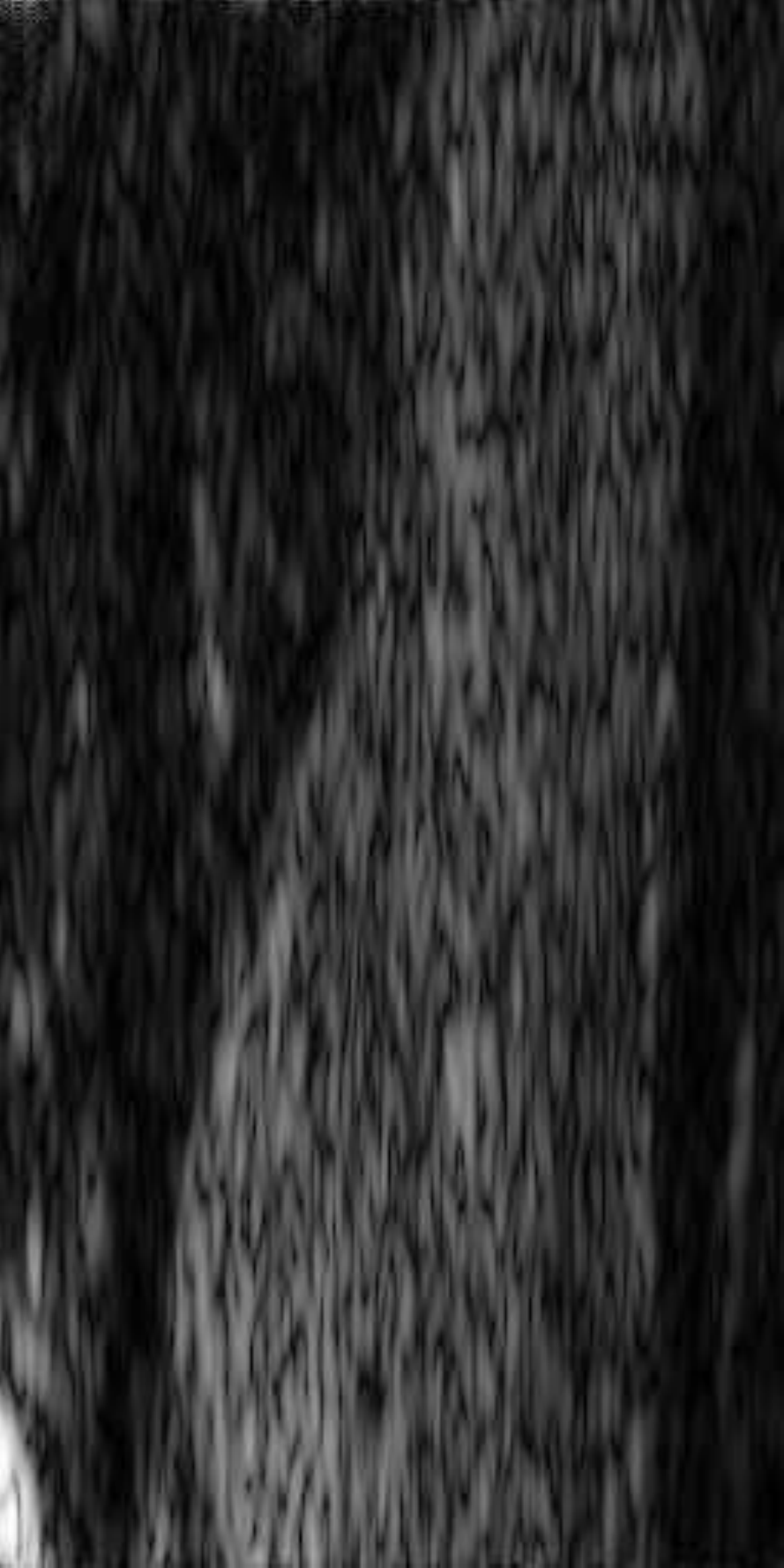}}\ \subfloat[]{\protect\includegraphics[width=1.75cm]{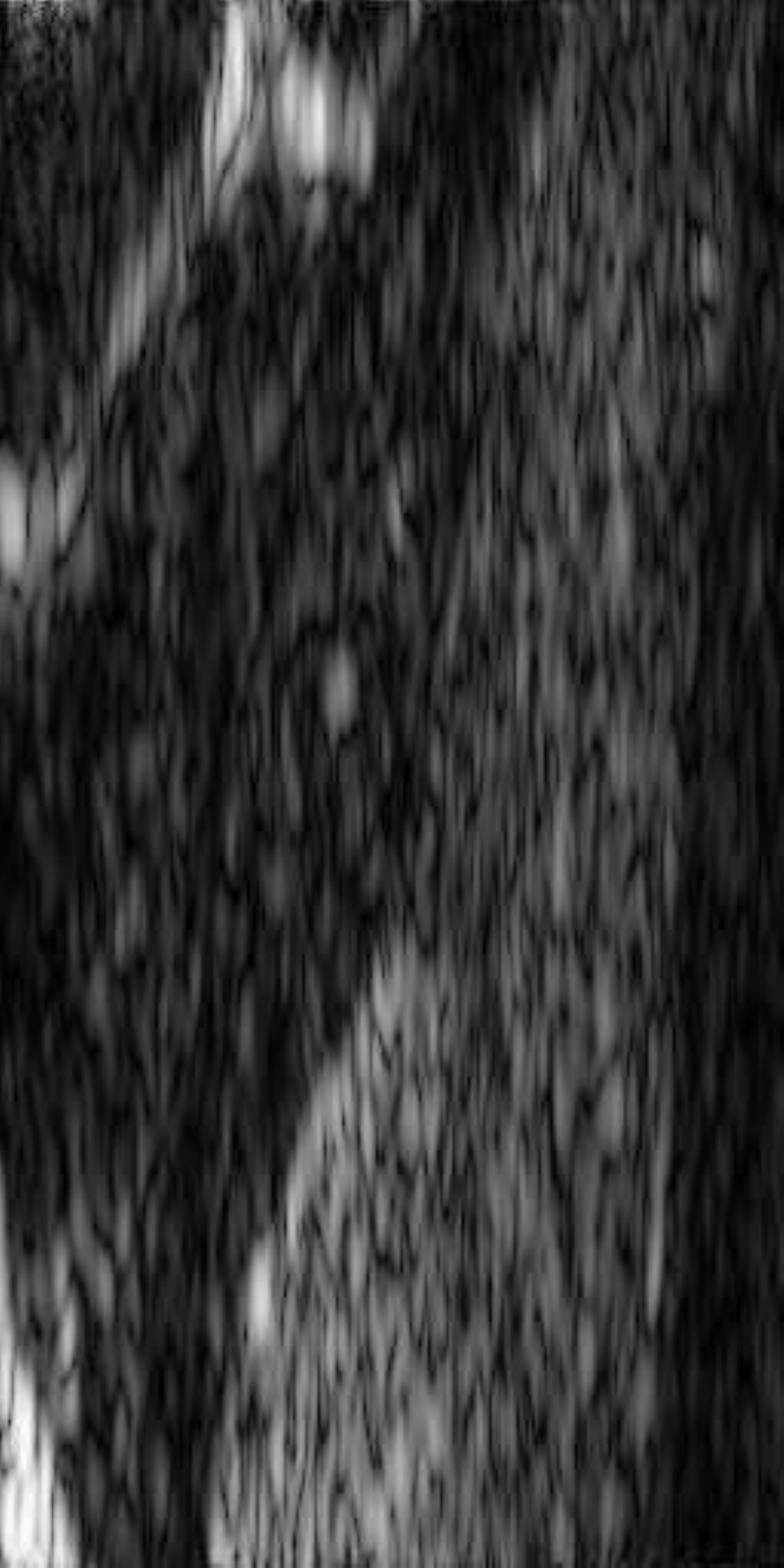}}\ \subfloat[]{\protect\includegraphics[width=1.75cm]{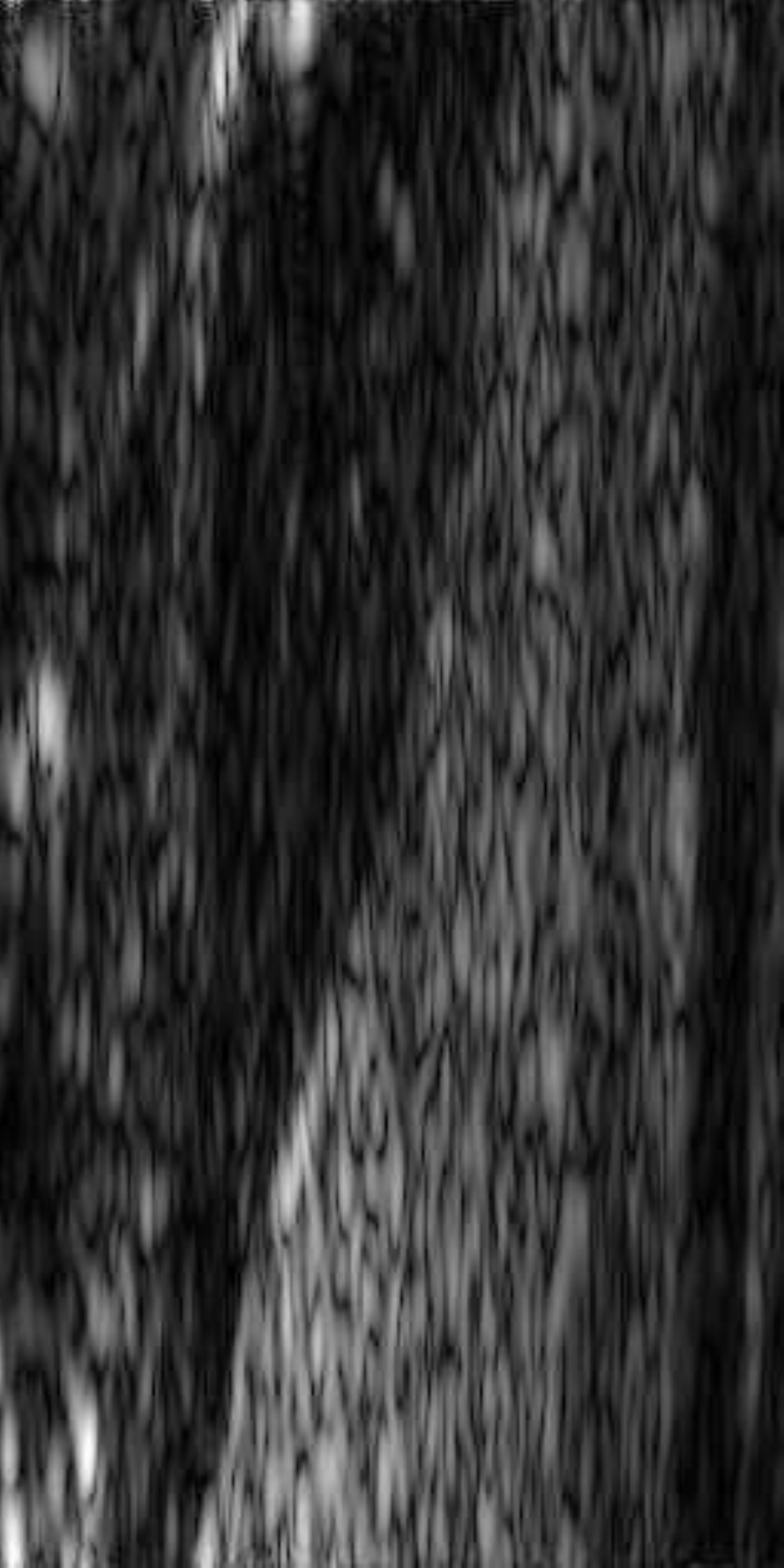}}
\par\end{centering}

\centering{}\subfloat[]{\protect\includegraphics[width=1.75cm]{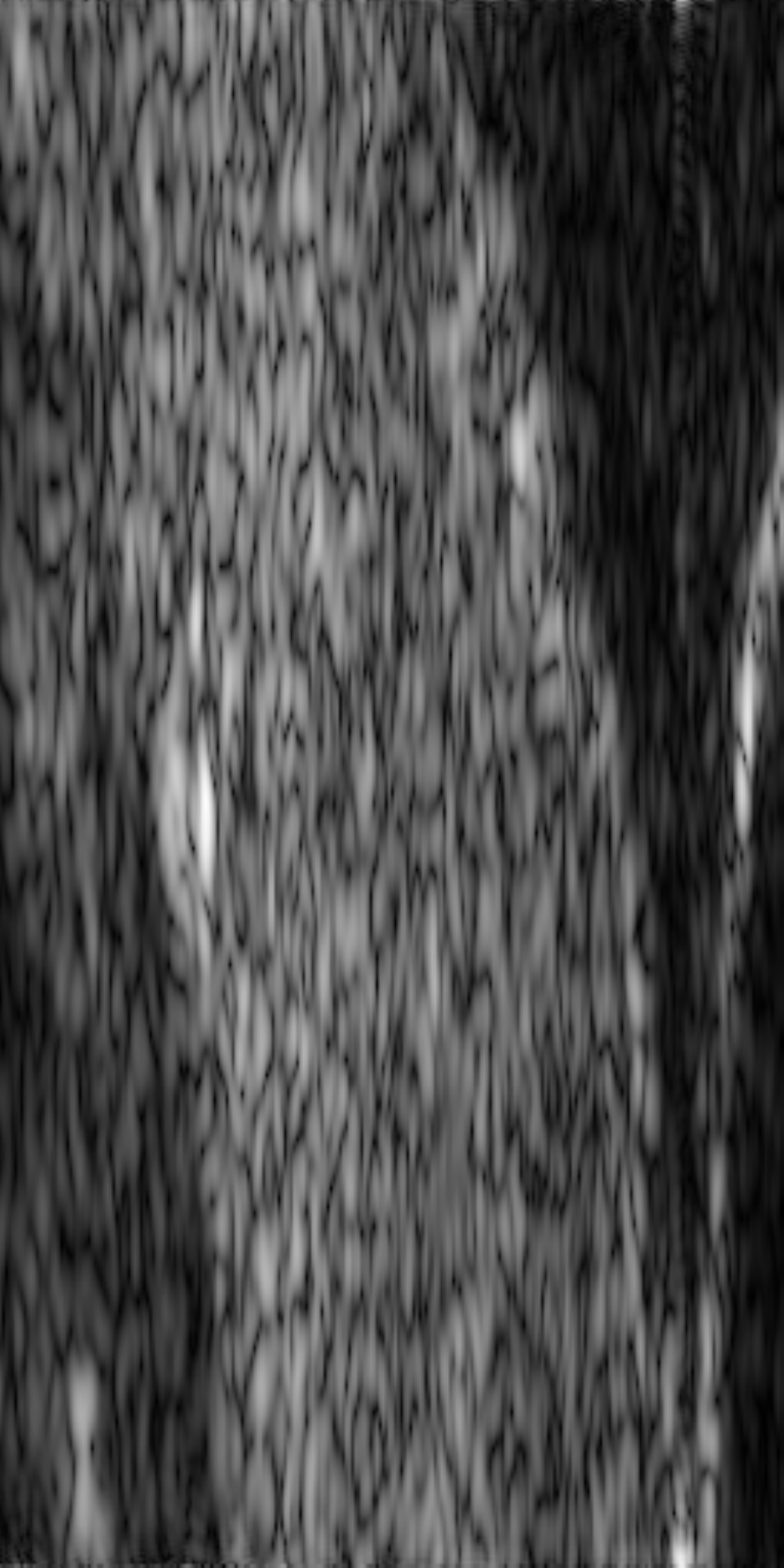}}\ \subfloat[]{\protect\includegraphics[width=1.75cm]{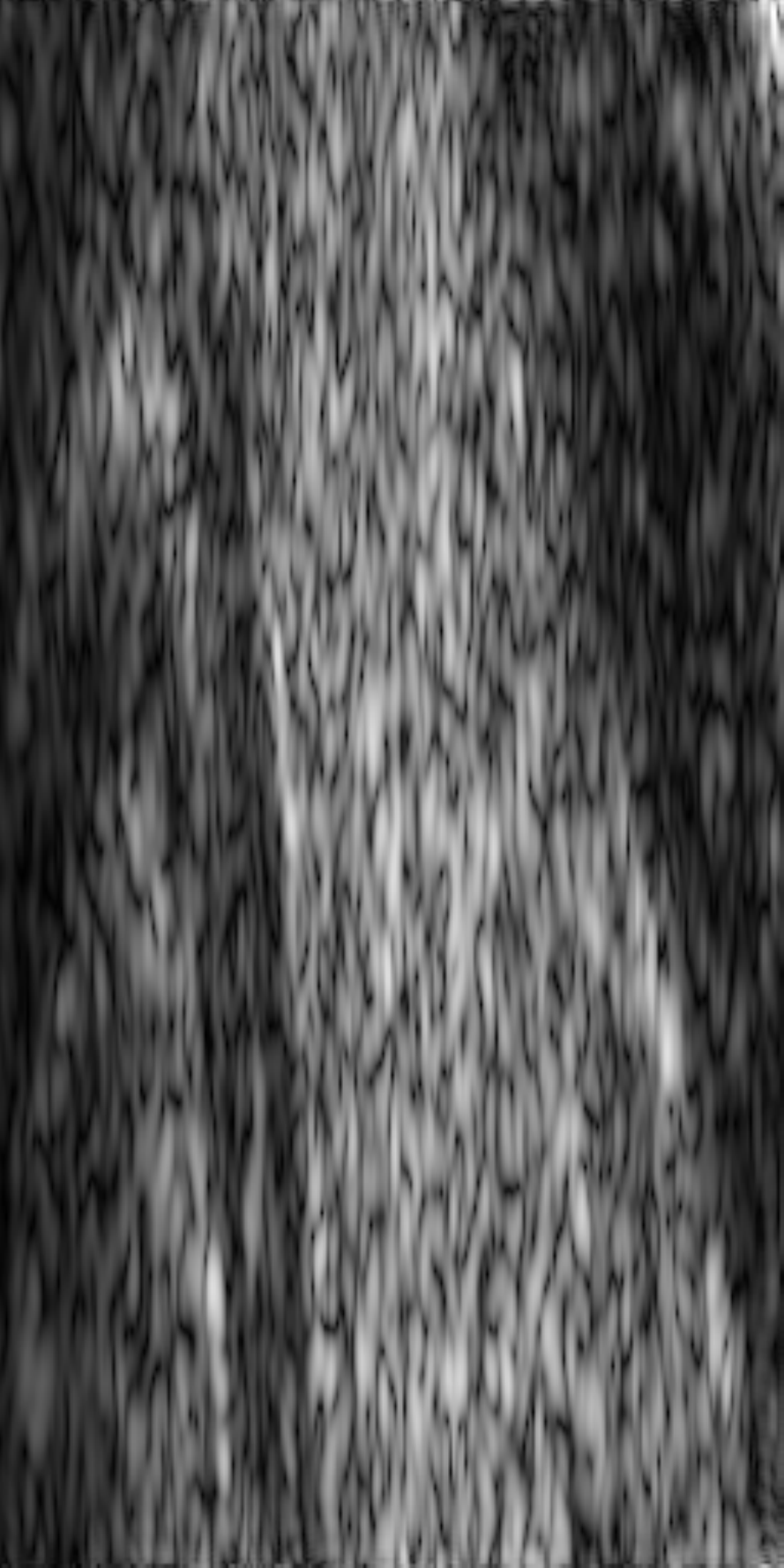}

}\ \subfloat[]{\protect\includegraphics[width=1.75cm]{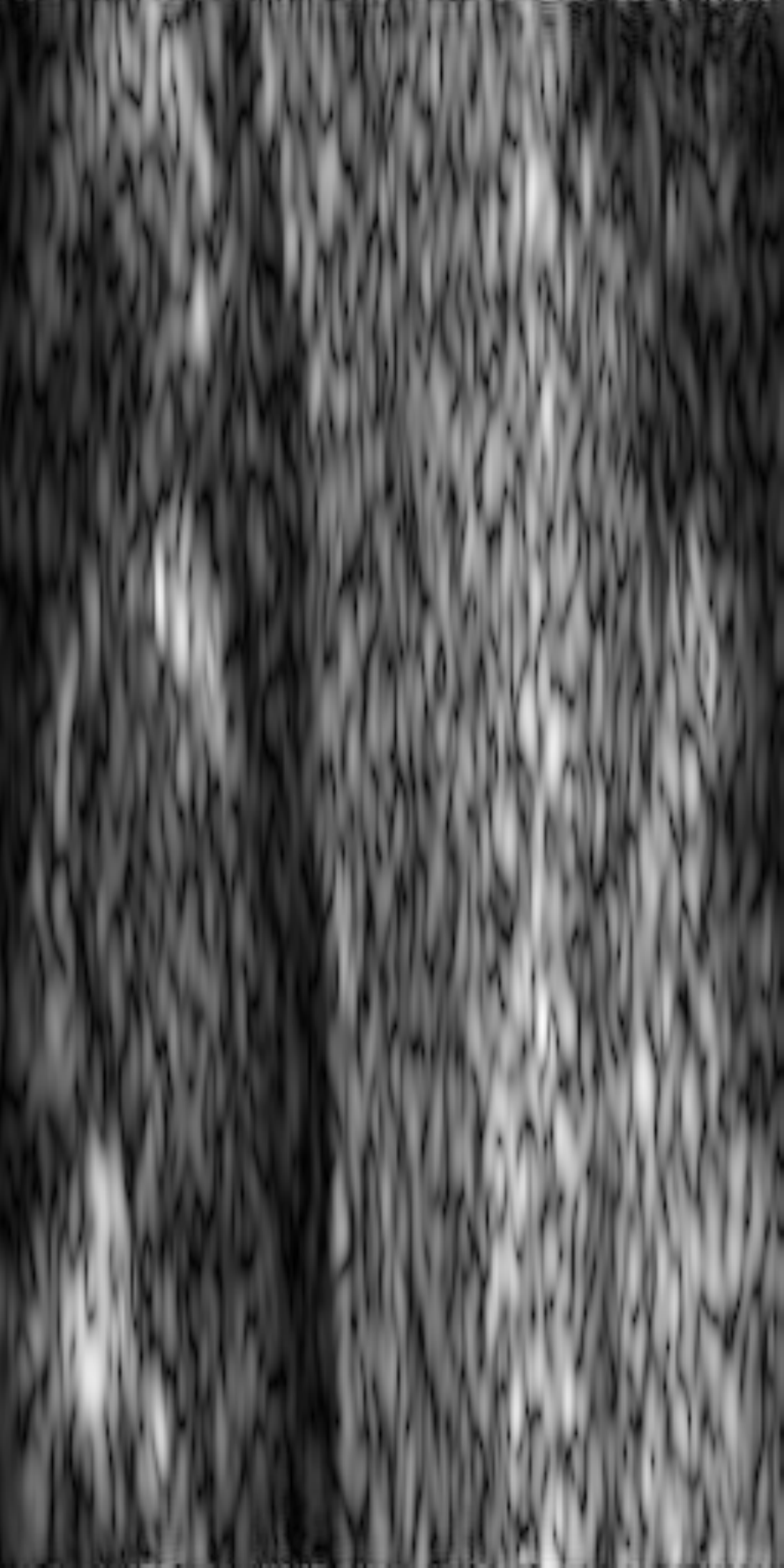}

}\protect\caption{Ultrasound images (a) 1, (b) 2, (c) 3, (d) 4, (e) 5 and (f) 6 from
set 1}
\end{figure}

The first set of images we use to test these methods come from the
same dataset as those used in \cite{achim2014reconstruction}. The
data corresponds to \textit{in vivo} healthy thyroid glands. The images
were acquired using a Siemens Sonoline Elegra scanner using a 7.5
MHz linear probe and a sampling frequency of 50MHz. Each of the 7
images we use for testing were acquired by cropping patches of size
$512\times256$ from the original images. That is, the images consist
of 256 lines, each of length 512.

We use a Discrete Cosine Transform (DCT) over a Discrete Fourier Transform
(DFT) to avoid mapping the original real data to complex data which
results in essentially doubling the amount of data as we would then
have effectively mapped the data from $\mathbb{R}^{M}$ to $\mathbb{C^{M}}$
before applying the sensing matrix to it, which is equivalent to a
mapping from $\mathbb{R}^{M}$ to $\mathbb{R}^{2M}$. It is possible
to take this into account by halving the sampling rate and then taking
into account the conjugate symmetry of a DFT of real data whilst recovering
the original signal, but this adds a layer of unnecessary complexity.
It is for similar reasons that the DCT is used in preference to the
DFT in several compression standards such as JPEG. Although it is
common to use 2D DCT or wavelet transforms for compressed sensing
of images, this relates to the use of a CCD to capture the images,
and as a CCD is a 2D array of sensors, this approach is useful. However,
for ultrasound, each ultrasound line corresponds to a single transducer,
and so a line by line approach provides a more practical approach.
Using a line-by-line approach also has the advantage that the reconstruction
of each line can be handled independently, and so parallelisation
is trivial. In addition, if we wish to compressively sense an $M\times M$
image, and we assume that the number of measurements we need is a
constant fraction of the number of pixels in the image, then compressively
sensing the entire image at once requires a sensing matrix with $\mathcal{O}(M^{4})$
entries, whereas the line by line approach requires a sensing matrix
with only $\mathcal{O}(M^{2})$ entries. Figure \ref{fig:DCT-of-image1-3}
shows the DCTs of image 1, 2, and 3 with logarithmic compression to
better highlight the locations of the non-zero elements. This shows
us that the assumption of sparsity in the frequency (DCT) domain is
reasonable and is therefore likely to lead to good results.

To simulate a compressive sampling system, we take the DCT of each
line of the image, and then project this onto a random Gaussian basis
at two levels, 33\% subsampling and 50\% subsampling, corresponding
to $\mathbf{A}\in\mathbb{R}^{171\times512}$ and $\mathbf{A}\in\mathbb{R}^{256\times512}$.

This simulation of the sensing system is as follows:
\begin{enumerate}
\item The DCT of each line (prior to envelope detection/logarithmic compression
being applied) of the original ultrasound data is calculated
\item Each of these DCTs is multiplied by the sensing matrix $\mathbf{A}$
\item The CS reconstruction methods are applied to recover these DCTs
\item The inverse DCT is then applied to each of the recovered lines
\item Envelope Detection/Logarithmic compression is then applied
\end{enumerate}
Note that the signals we sense are the raw ultrasound data and not
the displayed image. The displayed (B-mode) images are the images
we use to calculate the PSNR. These are created by taking the Hilbert
transform of the original data, adding this to the original data as
the complex component, taking the absolute value, logarithmically
compressing this and finally rescaling such that the smallest value
correspond to black (zero) and the largest value corresponds to white
(one).

\selectlanguage{british}%

\subsection{Brno dataset}

The final set of tests are on a set of 84 images from the Signal processing
laboratory of the Brno University of Technology. The description of
the dataset given is reproduced for reference below. 
\begin{quotation}
The database contains images of common carotid artery (CCA) of ten
volunteers (mean age 27.5 \textpm{} 3.5 years) with different weight
(mean weight 76.5 \textpm{} 9.7 kg). Images (usually eight images
per volunteer) were acquired with Sonix OP ultrasound scanner with
different set-up of depth, gain, time gain compensation (TGC) curve
and different linear array transducers. The image database contains
84 B-mode ultrasound images of CCA in longitudinal section. The resolution
of images is approximately 390x330px. The exact resolution depends
on the set-up of the ultrasound scanner. Two different linear array
transducers with different frequencies (10MHz and 14MHz) were used.
These frequencies were chosen because of their suitability for superficial
organs imaging. All images were taken by the specialists with five
year experience with scanning of arteries. Images were captured in
accordance to the standard protocol with patients lying in the supine
position and with the neck rotated to the left side while the right
CCA was examined.
\end{quotation}
It should be noted that these images, unlike those in the first data
set are provided after envelope detection and logarithmic compression,
and hence any differences in the results when compared to those in
the previous data set must be examined with this in mind. Before being
used to test the algorithms, the images were cropped such that the
height was a multiple of 32 and the width a multiple of 4. The simulated
sensing system works as it did in the previous section, except that
the envelope detection and logarithmic compression steps are skipped.

\selectlanguage{english}%

\section{Results}

\label{sec:results}

In this section, we conduct intensive tests in order to quantify the
performance of the various methods described in section \ref{sec:methods}.
In order to evaluate the performance of the algorithms, we calculate
the PSNR of the recovered images. The PSNR is given by equation (\ref{eq:-15-1}).

\begin{equation}
\text{PSNR}(\hat{I},I)=20\log_{10}(\text{MA\ensuremath{X_{I}}})-10\log_{10}(\text{MSE}(\hat{I},I)\label{eq:-15-1}
\end{equation}

\noindent Here, $\text{MAX}_{I}$ is the maximum possible pixel value
in the image.

In this section, ST-SBL $x/y$ refers to ST-SBL using a block size
of $y$ and processing columns in blocks of size $x$, and BSBL-BO
$x$ refers to BSBL-BO using a blocksize of $x$.

\subsection{Thyroid dataset}

We first consider the effect of block size on the performance of the
BSBL-BO method, fixing the pruning parameter $\bar{\Gamma}$ to be
$10^{-8}$. The block size can be thought of as the size we expect
clusters of non-zero elements in the DCT of each line of the ultrasound
image to be. We consider only the case where all block sizes are equal,
and therefore all block sizes we consider are powers of 2. In fact,
we tested all such block sizes, but present only the most relevant
results. In this case, we can see from Table \ref{tab:bsblboresults33108}
that at a subsampling rate of $33\%$, a block size of 32 provides
optimal recovery for all images, whereas for a subsampling rate of
$50\%$ we can see from Table \ref{tab:bsblboresults50108} that although
a block size of 32 is still optimal for most of the images, a block
size of $64$ now provides better results for some of the images.
This is slightly surprising, as we would expect the block structure
to be a property of the image being reconstructed, and not of the
sampling method.

We now consider the effect of block size for ST-SBL, along with considering
the effect of the number of columns processed at a time. As with block
size, we only consider processing the columns in blocks of equal size,
and therefore all column block sizes we consider are powers of 2.
As with BSBL-BO, we tested all such block sizes and column block sizes,
but present only the most relevant results. ST-SBL works on the assumption
of shared sparsity between columns, and we can therefore think of
the column block size as representing how fast we expect the sparsity
structure to change as we move between the DCTs of each line of the
ultrasound image, with smaller column block sizes corresponding to
faster changes. In this case, we can see from Table \ref{tab:stsblresults}
that at a 33\% subsampling rate, the optimal block size is typically
$32$, which is consistent with the results we obtained for BSBL-BO
and the optimal column block size is typically 1. For the images that
deviate from this, these parameters would still be close to optimum,
with a maximum loss in term of PSNR of $0.23$dB. Moving to a subsampling
rate of $50\%$, these parameters become optimal for all images.

Note that with $L=1$ ST-SBL reduces to the BSBL case, and so the
small difference observed between these methods is due to BSBL being
implemented with a bound optimization method and ST-SBL with an expectation-maximization
method, although there may also be other slight differences between
the implementations.

We now consider the effect of the pruning parameter $\bar{\Gamma}$.
This parameter controls when blocks are pruned, that is, at what level
blocks are assumed to be zero. It can be thought of as relating to
how small (in terms of the sum of squares of the block) we expect
a block to be before it no longer has a significant effect on the
quality of the recovered image.

For BSBL-BO, we can see from Table \ref{tab:bsblboresults} that decreasing
$\bar{\Gamma}$ to $2.22\times10^{-16}$ results at the 33\% subsampling
levels in a slight increase in performance, but does not change the
optimal block size. The pattern is repeated at the $50\%$ subsampling
level, with Table \ref{tab:bsblboresults50} showing a greater increase
in performance than at the 33\% subsampling level, but no change in
optimal block size.

For ST-SBL, decreasing $\bar{\Gamma}$ to $2.22\times10^{-16}$ results
at the 33\% subsampling level in no change in performance (results
are not shown as they are identical to those in Table \ref{tab:stsblresults}).
At a 50\% subsampling rate, Table \ref{tab:stsblresults50} shows
an increase in performance, but no change in optimal block size or
column block size. These results are consistent with results seen
with BSBL-BO.

\begin{table}[H]
\begin{centering}
\begin{tabular}{|c|c|c|c|}
\hline 
 & \multicolumn{3}{c|}{Block Size}\tabularnewline
\hline 
Image & 16 & 32 & 64\tabularnewline
\hline 
\hline 
1 & 43.46 & \textbf{43.96} & 41.86\tabularnewline
\hline 
2 & 35.59 & \textbf{36.57} & 33.76\tabularnewline
\hline 
3 & 35.65 & \textbf{35.86} & 31.91\tabularnewline
\hline 
4 & 39.48 & \textbf{39.86} & 37.00\tabularnewline
\hline 
5 & 37.04 & \textbf{37.41} & 34.80\tabularnewline
\hline 
6 & 38.76 & \textbf{39.11} & 36.25\tabularnewline
\hline 
7 & 35.01 & \textbf{43.44} & 40.95\tabularnewline
\hline 
\end{tabular}
\par\end{centering}

\protect\caption{\label{tab:bsblboresults33108}Results for BSBL-BO (PSNR) at a 33\%
subsampling rate ($\bar{\Gamma}=10^{-8}$)}
\end{table}

\begin{table}[H]
\begin{centering}
\begin{tabular}{|c|c|c|c|}
\hline 
 & \multicolumn{3}{c|}{Block Size}\tabularnewline
\hline 
Image & 16 & 32 & 64\tabularnewline
\hline 
\hline 
1 & 34.95 & 47.43 & \textbf{50.39}\tabularnewline
\hline 
2 & 42.42 & \textbf{44.16} & 35.00\tabularnewline
\hline 
3 & 42.65 & 31.52 & \textbf{42.90}\tabularnewline
\hline 
4 & 46.45 & \textbf{47.47} & 45.32\tabularnewline
\hline 
5 & 44.97 & \textbf{46.26} & 44.72\tabularnewline
\hline 
6 & 45.92 & \textbf{46.94} & 45.53\tabularnewline
\hline 
7 & 35.52 & 49.59 & \textbf{50.29}\tabularnewline
\hline 
\end{tabular}
\par\end{centering}

\protect\caption{\label{tab:bsblboresults50108}Results for BSBL-BO (PSNR) at a 50\%
subsampling rate ($\bar{\Gamma}=10^{-8}$)}
\end{table}

\begin{table}[H]
\centering{}%
\begin{tabular}{|c|c|c|c|c|c|c|}
\hline 
Column block size & \multicolumn{2}{c|}{1} & \multicolumn{2}{c|}{2} & \multicolumn{2}{c|}{4}\tabularnewline
\hline 
Block Size & 16 & 32 & 16 & 32 & 16 & 32\tabularnewline
\hline 
\hline 
Image 1 & 43.58 & \textbf{44.03} & 43.80 & 44.00 & 43.92 & 43.76\tabularnewline
\hline 
Image 2 & 35.56 & \textbf{36.62} & 35.80 & 36.28 & 34.47 & 35.11\tabularnewline
\hline 
Image 3 & \textbf{35.58} & 35.35 & 35.48 & 35.49 & 33.92 & 34.07\tabularnewline
\hline 
Image 4 & 39.62 & \textbf{39.94} & 39.94 & 39.72 & 39.81 & 39.34\tabularnewline
\hline 
Image 5 & 37.54 & 37.49 & \textbf{37.72} & 37.37 & 37.61 & 36.93\tabularnewline
\hline 
Image 6 & 39.00 & \textbf{39.34} & 39.14 & 39.34 & 39.04 & 38.96\tabularnewline
\hline 
Image 7 & 43.25 & 43.52 & 43.44 & 43.44 & \textbf{43.57} & 43.48\tabularnewline
\hline 
\end{tabular}\protect\caption{\label{tab:stsblresults}Results for ST-SBL (PSNR) at a 33\% subsampling
rate ($\bar{\Gamma}=10^{-8}$)}
\end{table}

\begin{table}[H]
\begin{tabular}{|c|c|c|c|c|c|c|}
\hline 
Column block size & \multicolumn{2}{c|}{1} & \multicolumn{2}{c|}{2} & \multicolumn{2}{c|}{4}\tabularnewline
\hline 
Block Size & 16 & 32 & 16 & 32 & 16 & 32\tabularnewline
\hline 
\hline 
Image 1 & 52.42 & \textbf{52.85} & 52.61 & 52.98 & 52.57 & 52.78\tabularnewline
\hline 
Image 2 & 44.22 & \textbf{45.91} & 43.29 & 44.55 & 43.72 & 43.79\tabularnewline
\hline 
Image 3 & 43.85 & \textbf{44.84} & 43.96 & 44.67 & 43.32 & 44.15\tabularnewline
\hline 
Image 4 & 48.92 & \textbf{49.77} & 49.08 & 49.55 & 49.00 & 49.17\tabularnewline
\hline 
Image 5 & 47.02 & \textbf{47.69} & 46.73 & 47.25 & 46.66 & 47.24\tabularnewline
\hline 
Image 6 & 48.91 & \textbf{49.59} & 48.55 & 49.08 & 48.73 & 48.84\tabularnewline
\hline 
Image 7 & 52.46 & \textbf{52.88} & 52.45 & 52.60 & 52.35 & 52.45\tabularnewline
\hline 
\end{tabular}\protect\caption{\label{tab:stsblresults50108}Results for ST-SBL (PSNR) at a 50\%
subsampling rate ($\bar{\Gamma}=10^{-8}$)}
\end{table}

\paragraph*{
\begin{table}[H]
\protect\centering{}%
\begin{tabular}{|c|c|c|}
\hline 
PSNR & \multicolumn{2}{c|}{Block Size}\tabularnewline
\hline 
Image & 16 & 32\tabularnewline
\hline 
\hline 
1 & 43.48 & \textbf{43.98}\tabularnewline
\hline 
2 & 35.67 & \textbf{36.49}\tabularnewline
\hline 
3 & 35.69 & \textbf{35.97}\tabularnewline
\hline 
4 & 39.64 & \textbf{39.93}\tabularnewline
\hline 
5 & 37.36 & \textbf{37.50}\tabularnewline
\hline 
6 & 38.88 & \textbf{39.24}\tabularnewline
\hline 
7 & 35.03 & \textbf{43.48}\tabularnewline
\hline 
\end{tabular}\protect\caption{\label{tab:bsblboresults}Results for BSBL-BO (PSNR) at a 33\% subsampling
rate ($\bar{\Gamma}=2.22\times10^{-16}$)}
\protect
\end{table}
}

\begin{table}[H]
\begin{centering}
\begin{tabular}{|c|c|c|c|}
\hline 
 & \multicolumn{3}{c|}{Block Size}\tabularnewline
\hline 
Image & 16 & 32 & 64\tabularnewline
\hline 
\hline 
1 & 34.96 & 45.69 & \textbf{52.71}\tabularnewline
\hline 
2 & 33.13 & \textbf{45.19} & 35.16\tabularnewline
\hline 
3 & 43.01 & 31.56 & \textbf{44.24}\tabularnewline
\hline 
4 & 49.18 & \textbf{50.01} & 49.23\tabularnewline
\hline 
5 & 47.06 & \textbf{48.05} & 47.60\tabularnewline
\hline 
6 & 48.87 & \textbf{49.45} & 49.00\tabularnewline
\hline 
7 & 35.42 & 50.49 & \textbf{52.87}\tabularnewline
\hline 
\end{tabular}
\par\end{centering}

\protect\caption{\label{tab:bsblboresults50}Results for BSBL-BO (PSNR) at a 50\% subsampling
rate ($\bar{\Gamma}=2.22\times10^{-16}$)}
\end{table}

\begin{table}[H]
\centering{}%
\begin{tabular}{|c|c|c|c|c|c|c|}
\hline 
Column block size & \multicolumn{2}{c|}{1} & \multicolumn{2}{c|}{2} & \multicolumn{2}{c|}{4}\tabularnewline
\hline 
Block Size & 16 & 32 & 16 & 32 & 16 & 32\tabularnewline
\hline 
\hline 
Image 1 & 52.75 & \textbf{53.27} & 52.75 & 53.13 & 52.69 & 52.91\tabularnewline
\hline 
Image 2 & 44.32 & \textbf{46.07} & 43.41 & 44.75 & 43.77 & 43.81\tabularnewline
\hline 
Image 3 & 43.89 & \textbf{44.90} & 43.98 & 44.69 & 43.33 & 44.16\tabularnewline
\hline 
Image 4 & 49.35 & \textbf{50.04} & 49.40 & 49.82 & 49.21 & 49.32\tabularnewline
\hline 
Image 5 & 47.15 & \textbf{47.89} & 46.84 & 47.28 & 46.75 & 47.32\tabularnewline
\hline 
Image 6 & 49.41 & \textbf{49.80} & 48.84 & 49.17 & 48.49 & 48.95\tabularnewline
\hline 
Image 7 & 52.65 & \textbf{53.09} & 52.48 & 52.67 & 52.40 & 52.56\tabularnewline
\hline 
\end{tabular}\protect\caption{\label{tab:stsblresults50}Results for ST-SBL (PSNR) at a 50\% subsampling
rate ($\bar{\Gamma}=2.22\times10^{-16}$)}
\end{table}

We now compare the results obtained with ST-SBL and BSBL-BO to the
other methods we described in section \ref{sec:methods}. Table \ref{tab:33Comparison-of-recovery}
shows comparisons with a number of other methods for recovery of compressively
sensed signals at a 33\% subsampling rate, and Table \ref{tab:50Comparison-of-recovery}
for a 50\% subsampling rate. Of the methods tested, IRLS dual prior
consistently has the worst performance. At a subsampling rate of 33\%,
T-MSBL-MoG-4 outperforms T-MSBL in 6 out of 7 cases, whereas when
we move to a 50\% subsampling rate, T-MSBL consistently outperforms
T-MSBL-MoG-4. In addition to these methods Tables \ref{tab:33Comparison-of-recovery}
and \ref{tab:50Comparison-of-recovery} also show the PSNR that would
be achieved by taking the 86 and 128 largest elements (in absolute
value) of the DCT of each ultrasound line respectively. 86 and 128
were chosen to be half the measurements taken, as all optimal methods
were SMV methods, and in this case, if the vector we wish to recover
is $k$-sparse, a minimum of $2k$ measurements are required. Interestingly,
both ST-SBL 1/32 ($\bar{\Gamma}=2.22\times10^{-16}$) and BSBL-BO
($\bar{\Gamma}=2.22\times10^{-16}$) performed significantly better
than this method, suggesting that methods seeking to approximate the
$k$-sparse approximation may not be ideal.

\begin{table}[H]
\begin{centering}
\begin{tabular}{|c|>{\centering}p{2cm}|>{\centering}p{2cm}|>{\centering}p{1.7cm}|}
\hline 
 & \multicolumn{3}{c|}{Method}\tabularnewline
\hline 
{\footnotesize{}Image} & {\footnotesize{}ST-SBL 1/32 ($\bar{\Gamma}=2.22\times10^{-16}$)} & {\footnotesize{}BSBL-BO 32 ($\bar{\Gamma}=2.22\times10^{-16}$)} & {\footnotesize{}IRLS - Dual prior}\tabularnewline
\hline 
\hline 
{\footnotesize{}1} & {\footnotesize{}44.03} & {\footnotesize{}43.98} & {\footnotesize{}30.01}\tabularnewline
\hline 
{\footnotesize{}2} & {\footnotesize{}36.62} & {\footnotesize{}36.49} & {\footnotesize{}26.12}\tabularnewline
\hline 
{\footnotesize{}3} & {\footnotesize{}35.35} & {\footnotesize{}35.97} & {\footnotesize{}27.54}\tabularnewline
\hline 
{\footnotesize{}4} & {\footnotesize{}39.94} & {\footnotesize{}39.93} & {\footnotesize{}28.74}\tabularnewline
\hline 
{\footnotesize{}5} & {\footnotesize{}37.49} & {\footnotesize{}37.50} & {\footnotesize{}28.71}\tabularnewline
\hline 
{\footnotesize{}6} & {\footnotesize{}39.34} & {\footnotesize{}39.24} & {\footnotesize{}30.09}\tabularnewline
\hline 
{\footnotesize{}7} & {\footnotesize{}43.52} & {\footnotesize{}43.48} & {\footnotesize{}25.34}\tabularnewline
\hline 
\end{tabular}\\

\par\end{centering}

\begin{centering}
\begin{tabular}{|c|>{\centering}p{2cm}|>{\centering}p{2cm}|>{\centering}p{1.7cm}|}
\hline 
 & \multicolumn{3}{c|}{Method}\tabularnewline
\hline 
{\footnotesize{}Image} & {\footnotesize{}T-MSBL} & {\footnotesize{}T-MSBL-MoG-4} & {\footnotesize{}86-spase}\tabularnewline
\hline 
\hline 
{\footnotesize{}1} & {\footnotesize{}31.11} & {\footnotesize{}31.53} & {\footnotesize{}38.11}\tabularnewline
\hline 
{\footnotesize{}2} & {\footnotesize{}26.92} & {\footnotesize{}26.94} & {\footnotesize{}34.56}\tabularnewline
\hline 
{\footnotesize{}3} & {\footnotesize{}28.34} & {\footnotesize{}29.78} & {\footnotesize{}35.91}\tabularnewline
\hline 
{\footnotesize{}4} & {\footnotesize{}29.78} & {\footnotesize{}29.31} & {\footnotesize{}36.10}\tabularnewline
\hline 
{\footnotesize{}5} & {\footnotesize{}29.92} & {\footnotesize{}30.75} & {\footnotesize{}34.58}\tabularnewline
\hline 
{\footnotesize{}6} & {\footnotesize{}31.66} & {\footnotesize{}32.28} & {\footnotesize{}36.72}\tabularnewline
\hline 
{\footnotesize{}7} & {\footnotesize{}29.05} & {\footnotesize{}27.64} & {\footnotesize{}36.58}\tabularnewline
\hline 
\end{tabular}
\par\end{centering}

\protect\caption{\label{tab:33Comparison-of-recovery}Comparison of recovery results
at 33\% subsampling level (PSNR)}
\end{table}

\begin{table}[H]
\begin{centering}
{\footnotesize{}}%
\begin{tabular}{|c|>{\centering}p{2cm}|>{\centering}p{2cm}|>{\centering}p{1.7cm}|}
\hline 
 & \multicolumn{3}{c|}{{\footnotesize{}Method}}\tabularnewline
\hline 
{\footnotesize{}Image} & {\footnotesize{}ST-SBL 1/32 ($\bar{\Gamma}=2.22\times10^{-16}$)} & {\footnotesize{}BSBL-BO 32 ($\bar{\Gamma}=2.22\times10^{-16}$)} & {\footnotesize{}IRLS - Dual prior}\tabularnewline
\hline 
\hline 
{\footnotesize{}1} & {\footnotesize{}53.27} & {\footnotesize{}45.69} & {\footnotesize{}33.43}\tabularnewline
\hline 
{\footnotesize{}2} & {\footnotesize{}46.07} & {\footnotesize{}45.19} & {\footnotesize{}29.64}\tabularnewline
\hline 
{\footnotesize{}3} & {\footnotesize{}44.90} & {\footnotesize{}31.56} & {\footnotesize{}31.29}\tabularnewline
\hline 
{\footnotesize{}4} & {\footnotesize{}50.04} & {\footnotesize{}50.01} & {\footnotesize{}32.78}\tabularnewline
\hline 
{\footnotesize{}5} & {\footnotesize{}47.89} & {\footnotesize{}48.05} & {\footnotesize{}32.23}\tabularnewline
\hline 
{\footnotesize{}6} & {\footnotesize{}49.80} & {\footnotesize{}49.45} & {\footnotesize{}33.13}\tabularnewline
\hline 
{\footnotesize{}7} & {\footnotesize{}53.09} & {\footnotesize{}50.49} & {\footnotesize{}32.49}\tabularnewline
\hline 
\end{tabular}{\footnotesize{}}\\

\par\end{centering}{\footnotesize \par}

\begin{centering}
{\footnotesize{}}%
\begin{tabular}{|c|>{\centering}p{2cm}|>{\centering}p{2cm}|>{\centering}p{1.7cm}|}
\hline 
 & \multicolumn{3}{c|}{Method}\tabularnewline
\hline 
{\footnotesize{}Image} & {\footnotesize{}T-MSBL} & {\footnotesize{}T-MSBL-MoG-4} & {\footnotesize{}128-sparse}\tabularnewline
\hline 
\hline 
{\footnotesize{}1} & {\footnotesize{}37.44} & {\footnotesize{}37.31} & {\footnotesize{}41.80}\tabularnewline
\hline 
{\footnotesize{}2} & {\footnotesize{}32.62} & {\footnotesize{}30.37} & {\footnotesize{}38.15}\tabularnewline
\hline 
{\footnotesize{}3} & {\footnotesize{}35.25} & {\footnotesize{}34.94} & {\footnotesize{}39.26}\tabularnewline
\hline 
{\footnotesize{}4} & {\footnotesize{}34.02} & {\footnotesize{}34.01} & {\footnotesize{}39.78}\tabularnewline
\hline 
{\footnotesize{}5} & {\footnotesize{}35.87} & {\footnotesize{}35.84} & {\footnotesize{}38.69}\tabularnewline
\hline 
{\footnotesize{}6} & {\footnotesize{}38.10} & {\footnotesize{}37.70} & {\footnotesize{}40.82}\tabularnewline
\hline 
{\footnotesize{}7} & {\footnotesize{}35.27} & {\footnotesize{}35.12} & {\footnotesize{}39.73}\tabularnewline
\hline 
\end{tabular}
\par\end{centering}{\footnotesize \par}

\protect\caption{\label{tab:50Comparison-of-recovery}Comparison of recovery results
at 50\% subsampling level (PSNR)}
\end{table}

\begin{figure}[H]
\subfloat[]{\protect\includegraphics[width=1.75cm]{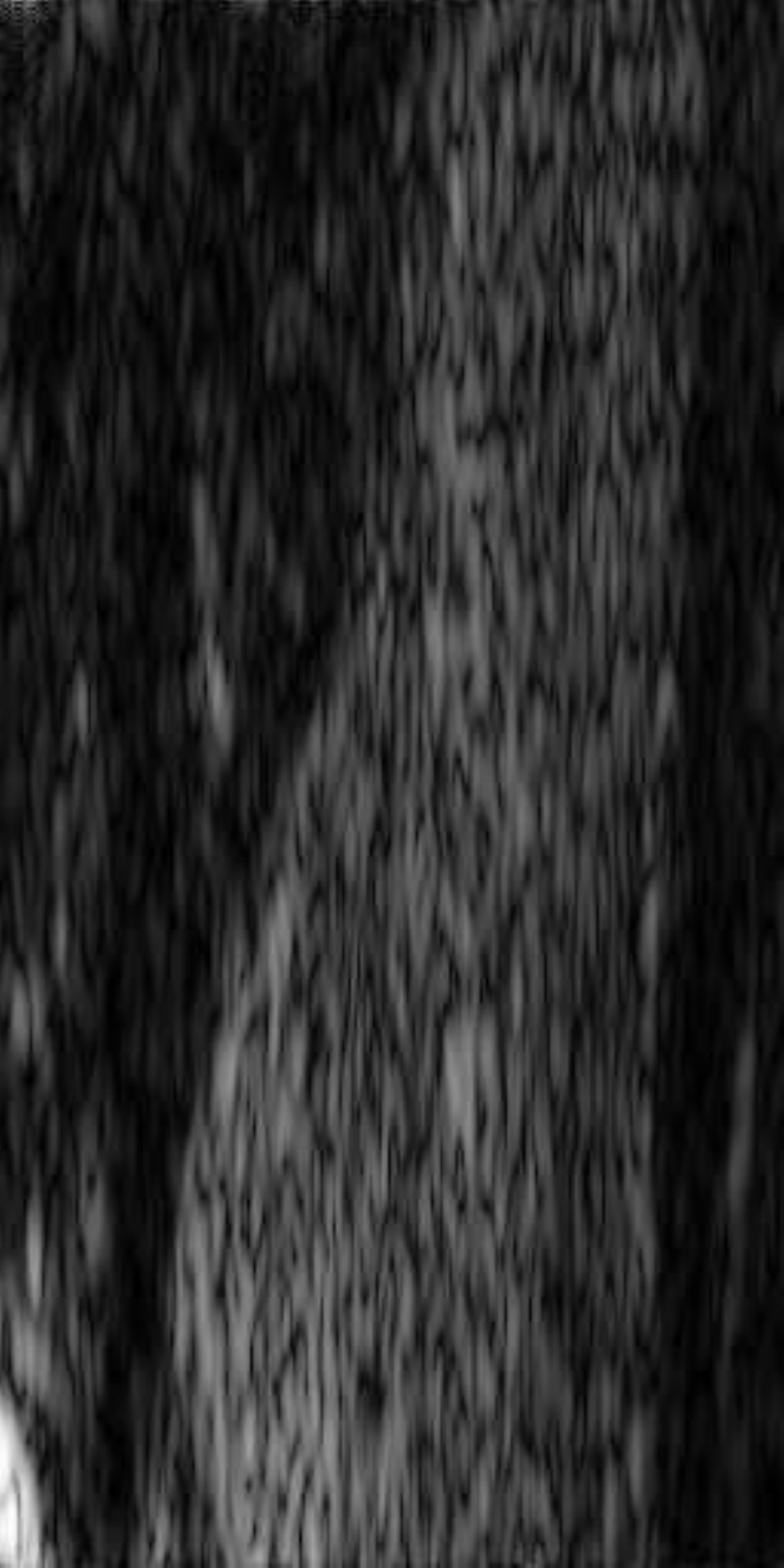}

}\hfill{} \subfloat[]{\protect\includegraphics[width=1.75cm]{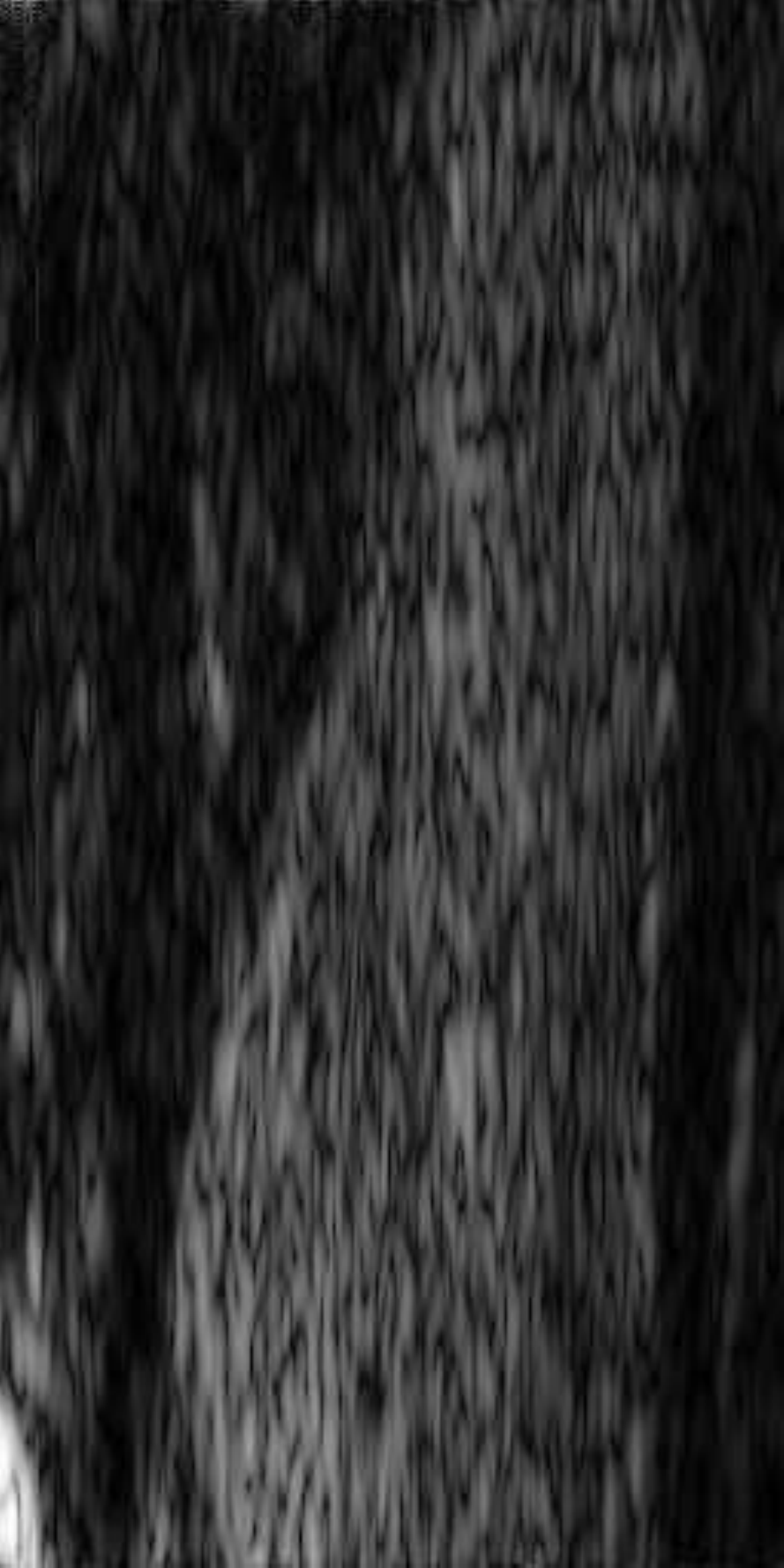}}\hfill{}
\subfloat[]{\protect\includegraphics[width=1.75cm]{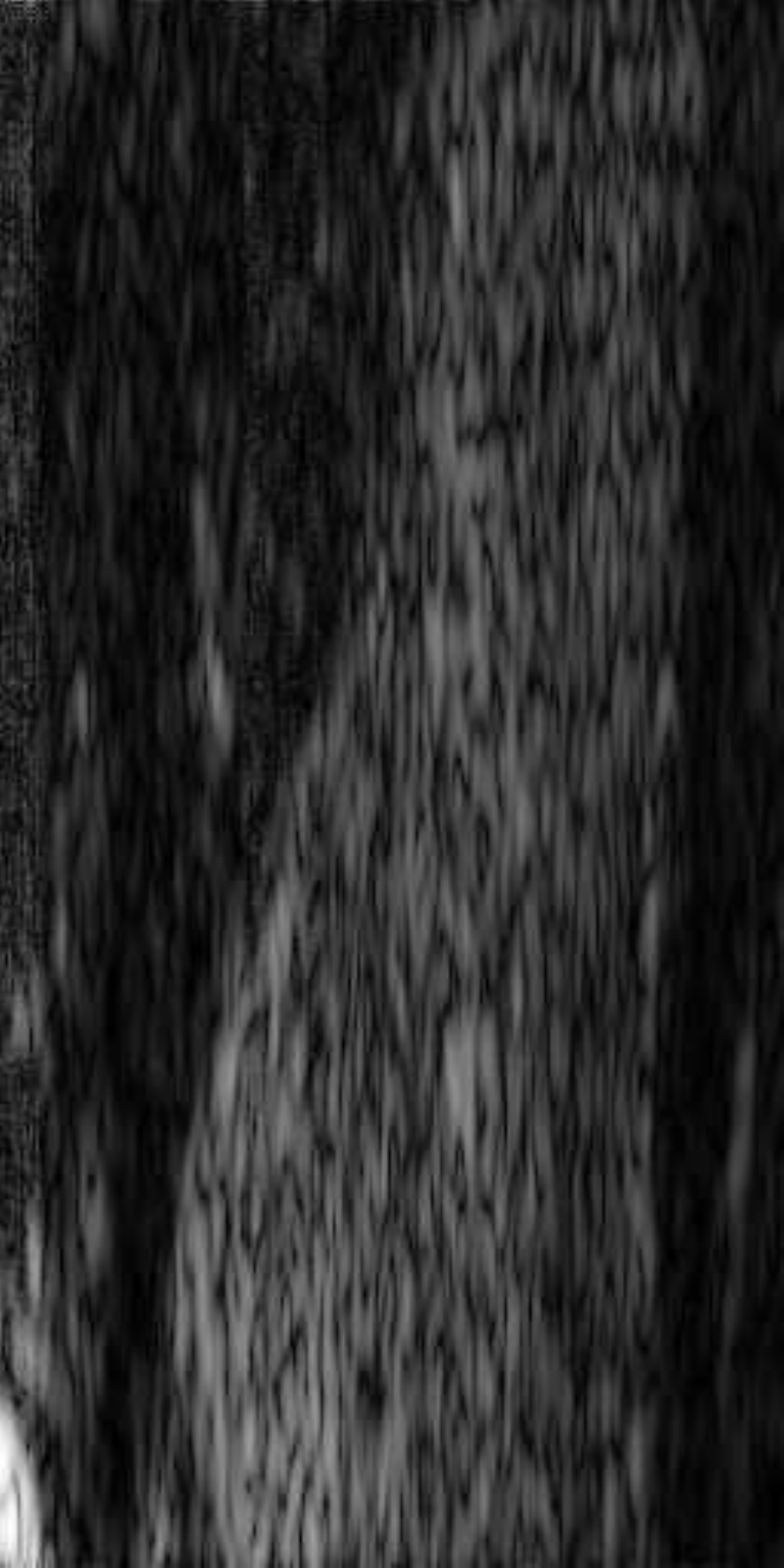}}\\
\subfloat[]{\protect\includegraphics[width=1.75cm]{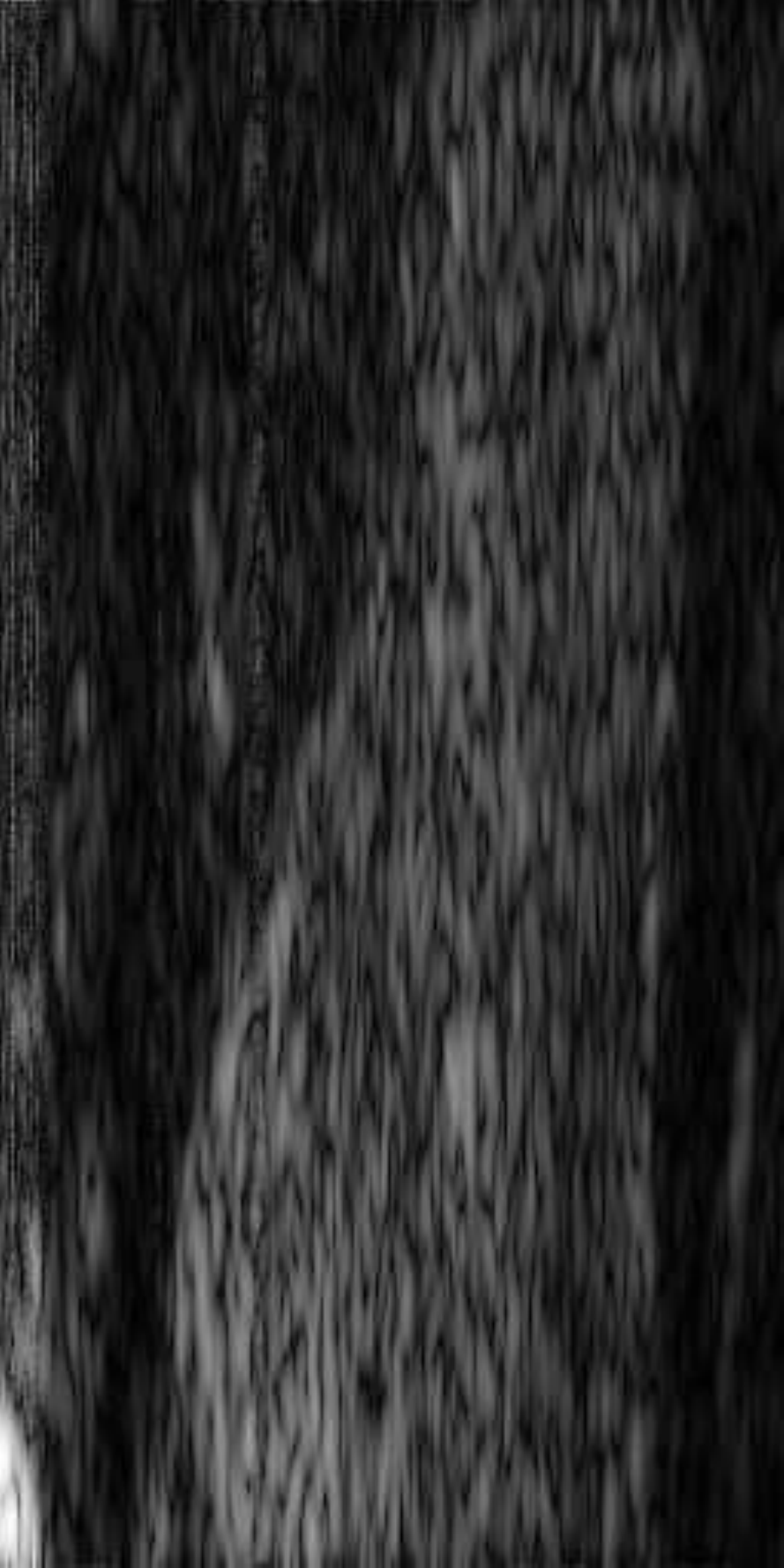}}
\hfill{}\subfloat[]{\protect\includegraphics[width=1.75cm]{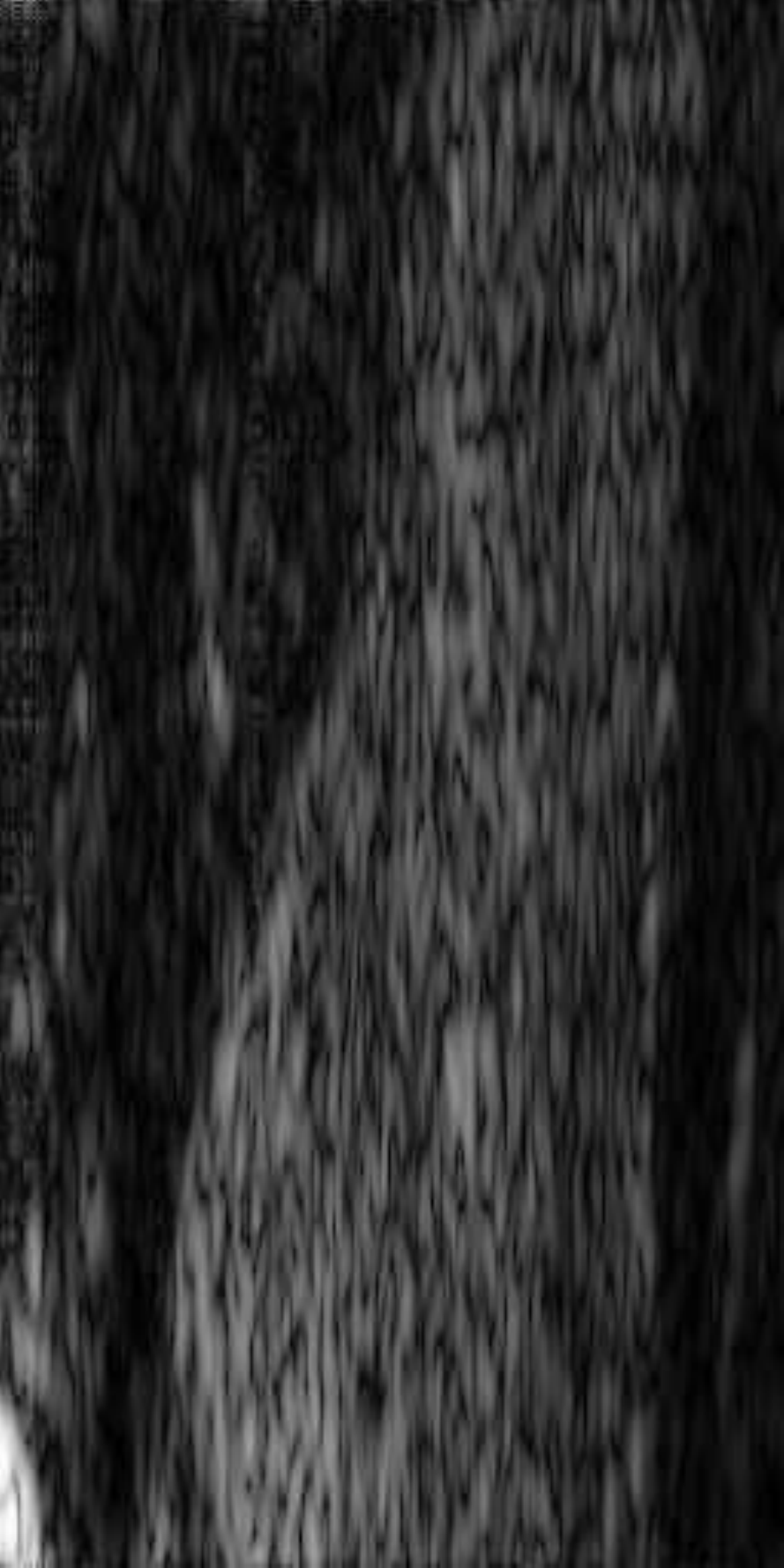}}\hfill{}
\subfloat[]{\protect\includegraphics[width=1.75cm]{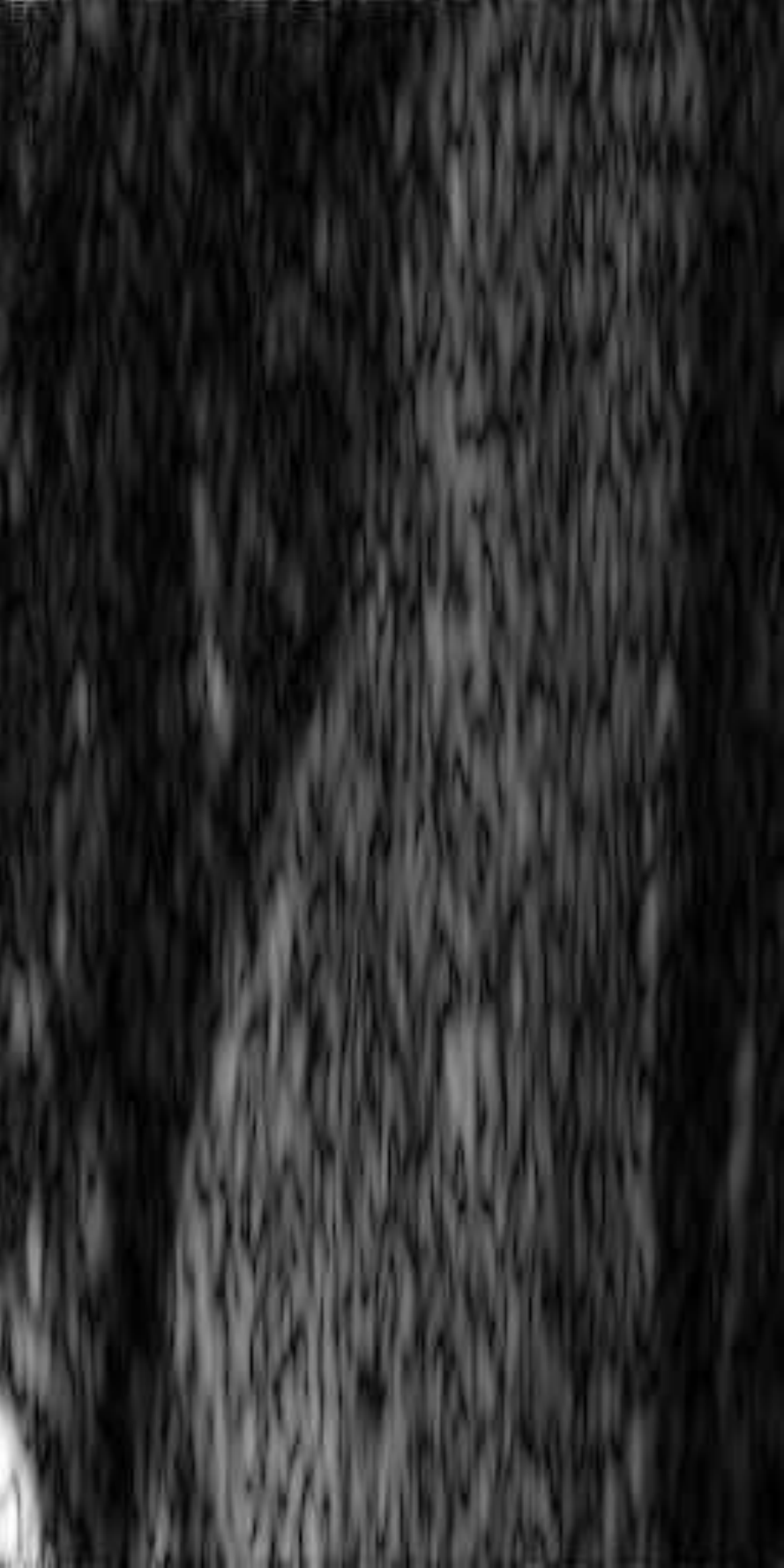}}

\protect\caption{\label{fig:Image150results}Image 1, subsampled at 50\% and recovered
with (a) ST-SBL 1/32 ($\bar{\Gamma}=2.22\times10^{-16}$), (b) BSBL-BO
32 ($\bar{\Gamma}=2.22\times10^{-16}$), (c) IRLS - Dual prior, (d)
T-MSBL, (e) T-MSBL-MoG-4, (f) 128-sparse}
\end{figure}

\begin{figure}[H]
\subfloat[]{\protect\includegraphics[width=1.75cm]{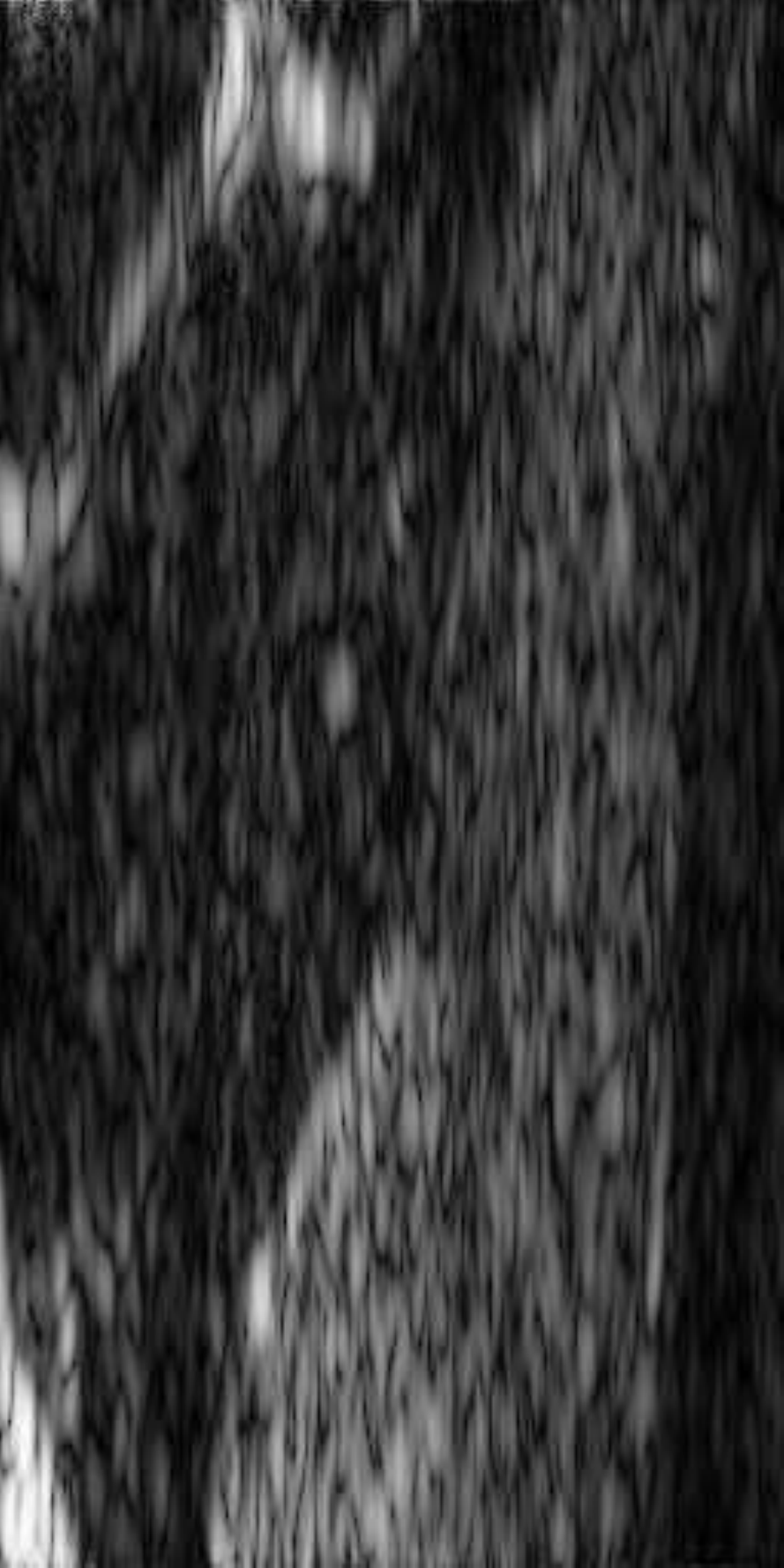}}\hfill{}\subfloat[]{\protect\includegraphics[width=1.75cm]{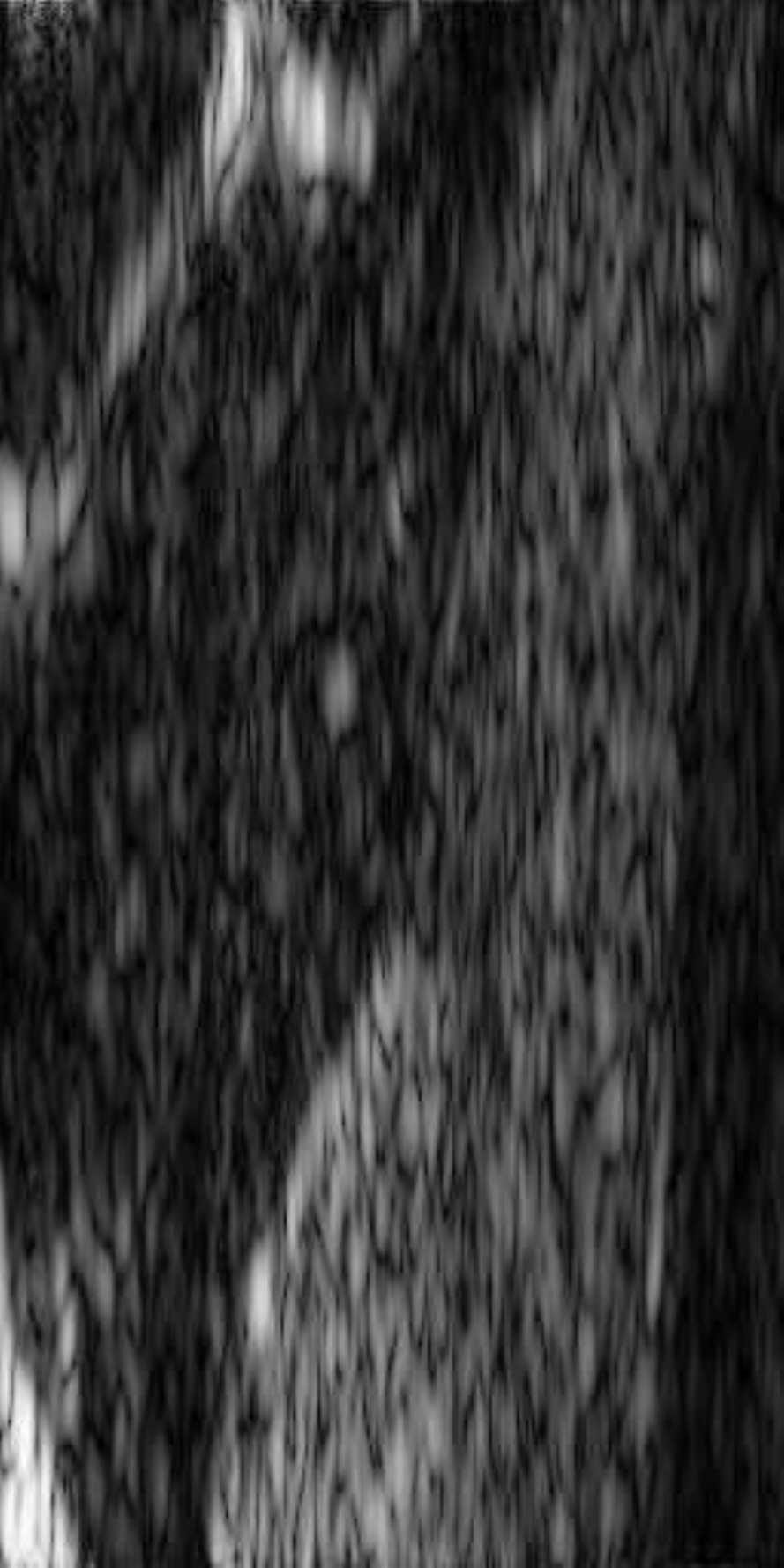}}\hfill{}\subfloat[]{\protect\includegraphics[width=1.75cm]{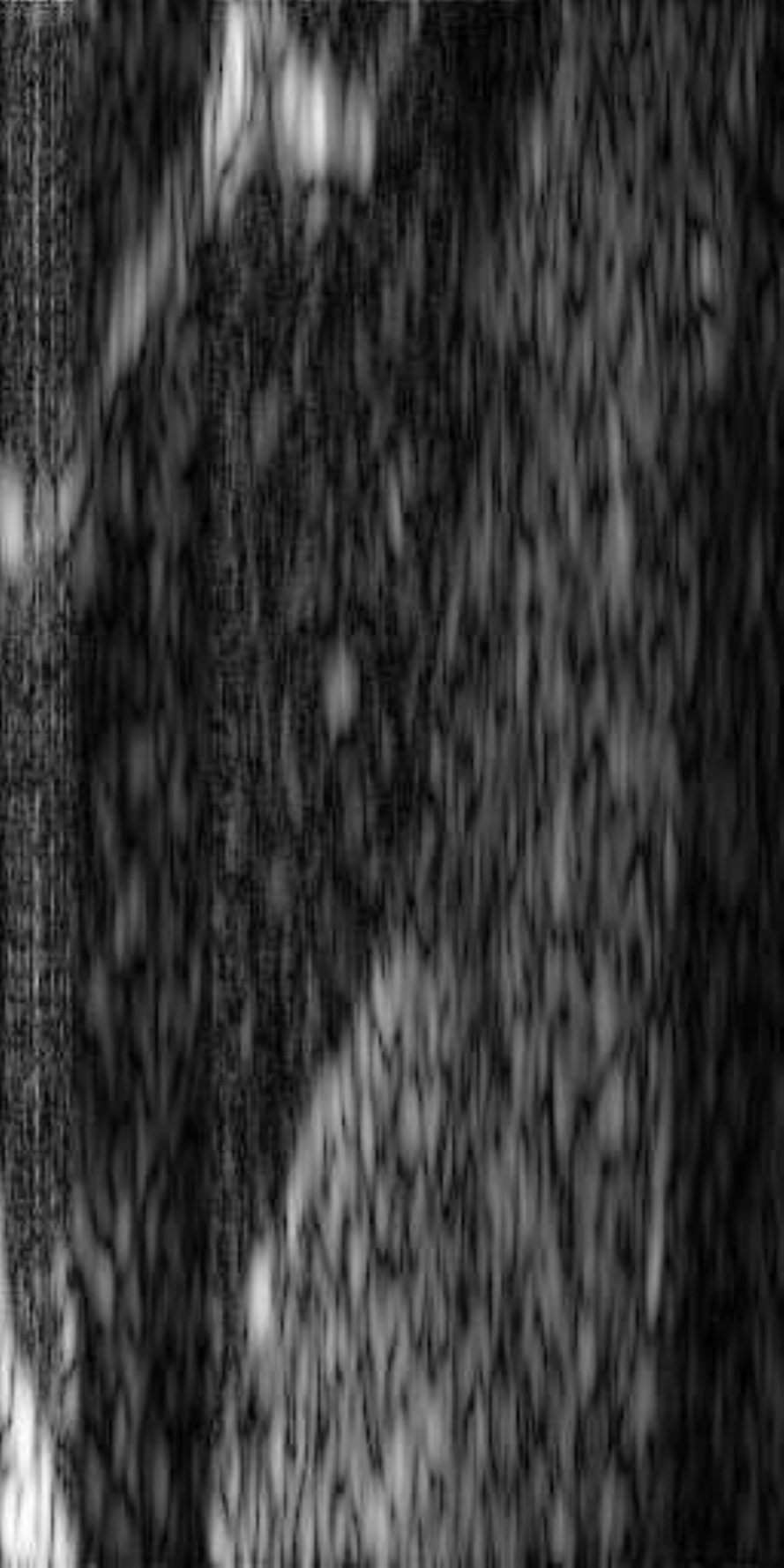}}\\
\subfloat[]{\protect\includegraphics[width=1.75cm]{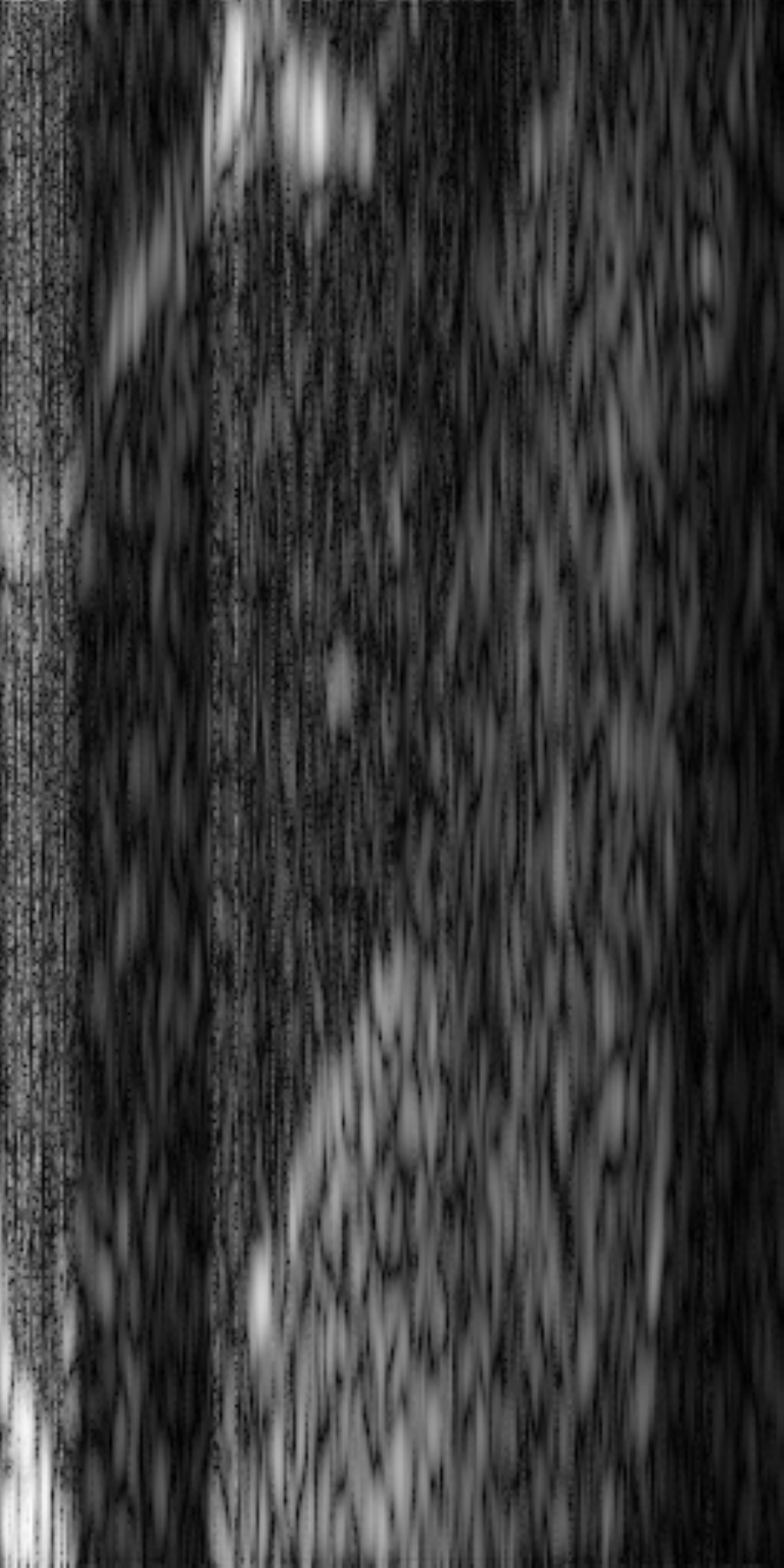}}\hfill{}\subfloat[]{\protect\includegraphics[width=1.75cm]{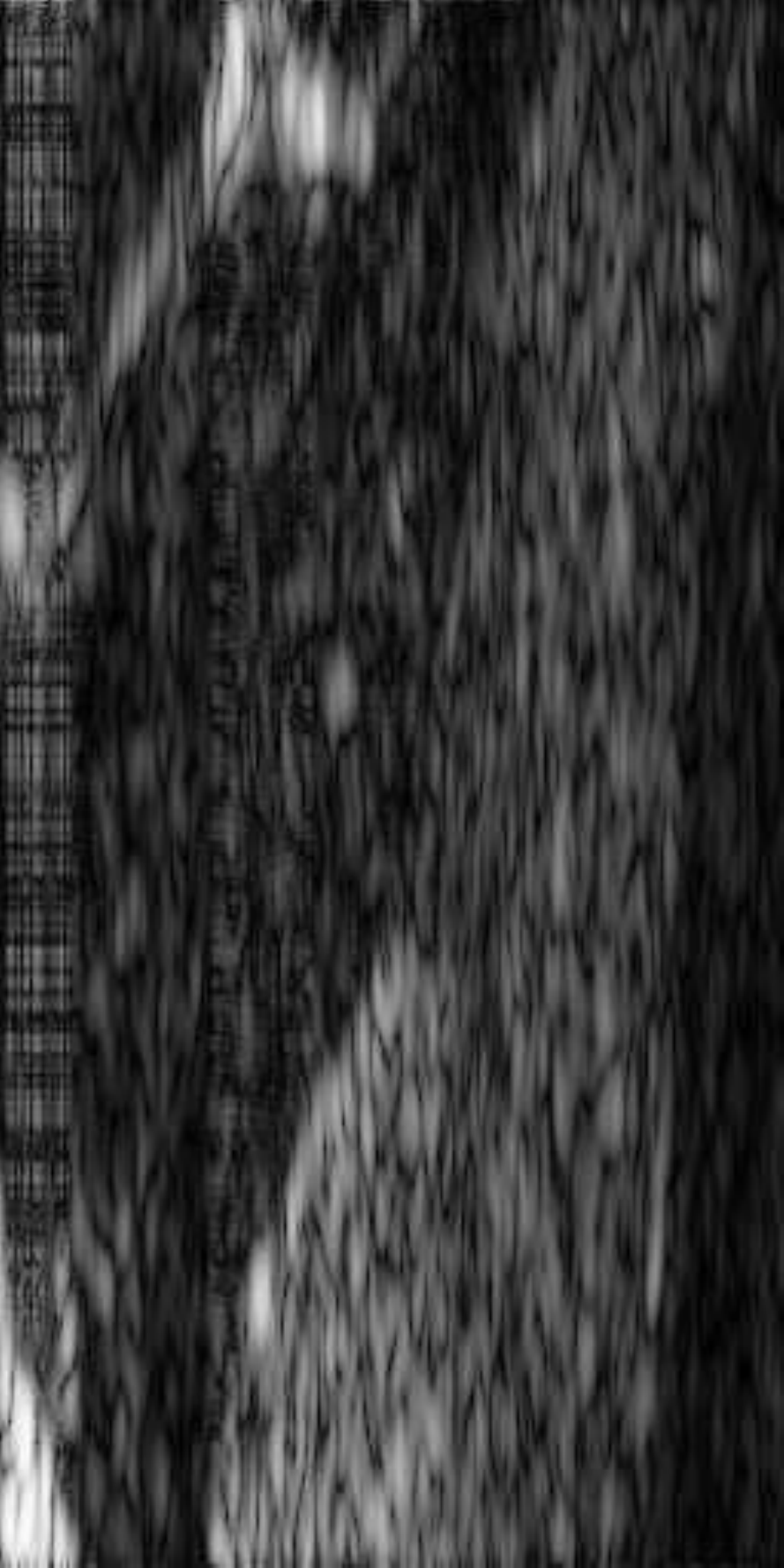}}\hfill{}\subfloat[]{\protect\includegraphics[width=1.75cm]{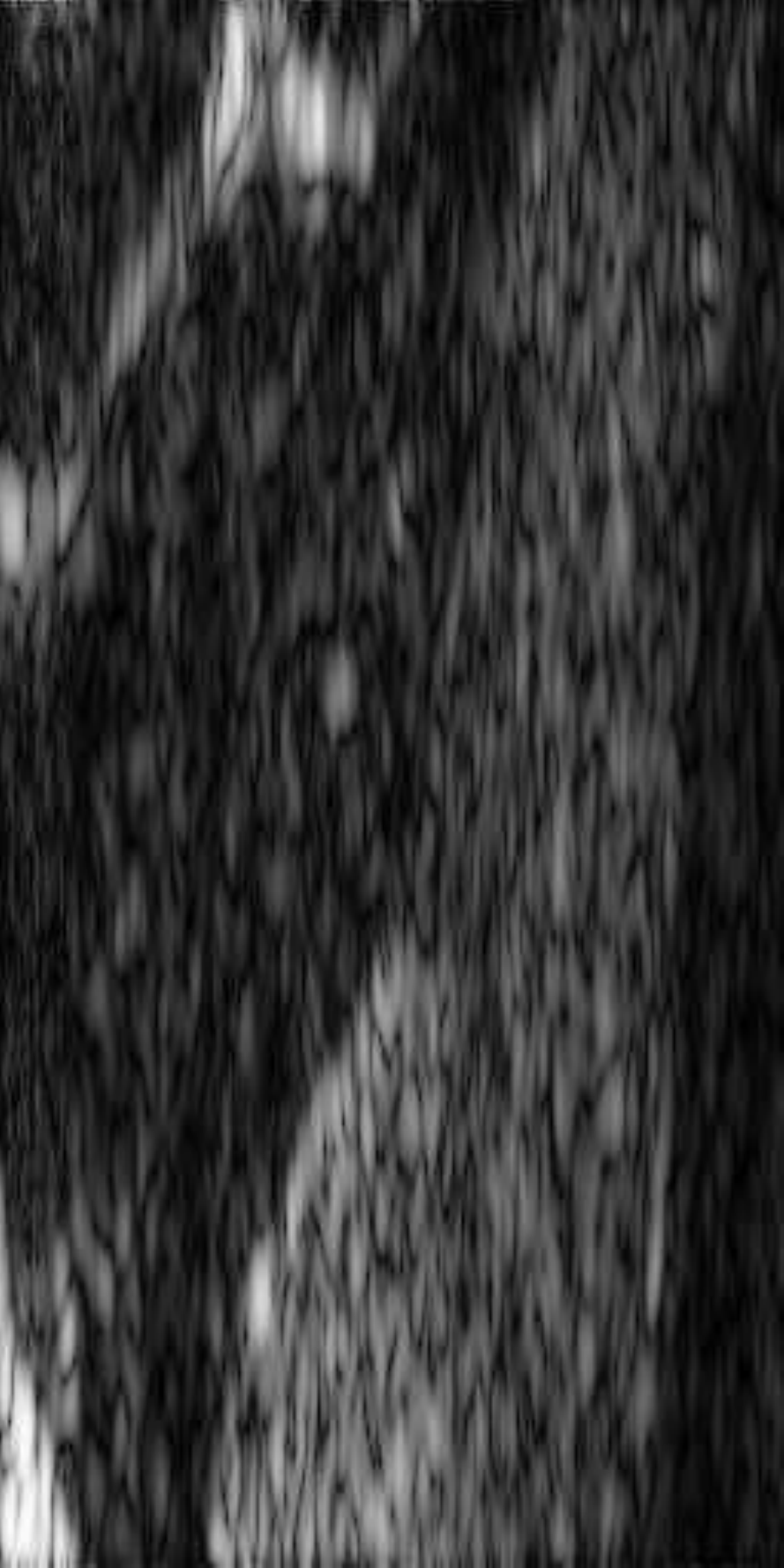}}

\protect\caption{\label{fig:Image233results}Image 2, subsampled at 33\% and recovered
with (a) ST-SBL 1/32 ($\bar{\Gamma}=2.22\times10^{-16}$), (b) BSBL-BO
32 ($\bar{\Gamma}=2.22\times10^{-16}$), (c) IRLS - Dual prior, (d)
T-MSBL, (e) T-MSBL-MoG-4, (f) 86-sparse}
\end{figure}

\begin{figure}[tbh]
\includegraphics[width=8.5cm]{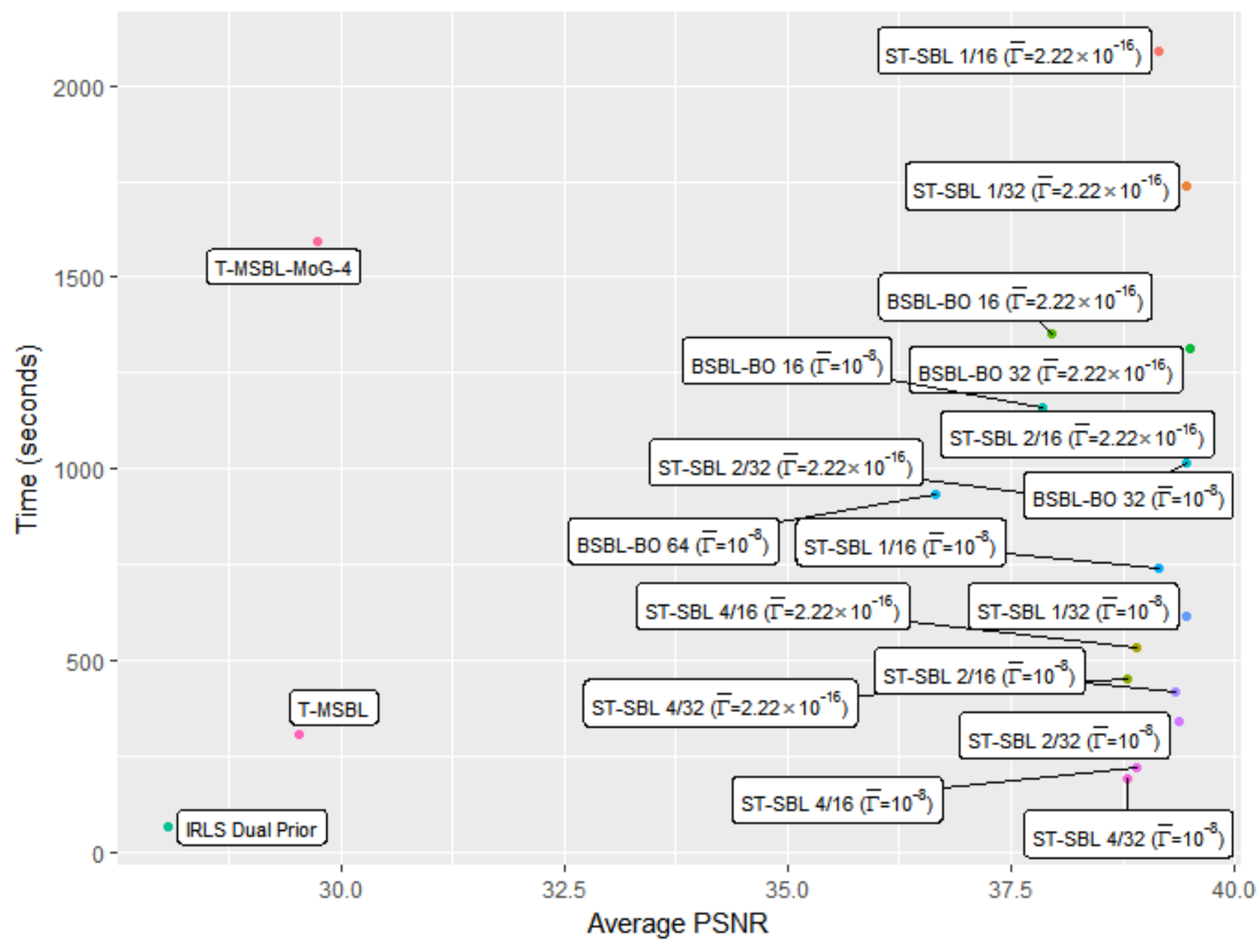}

\protect\caption{\label{fig:Time-vs-Performance33}Time vs Performance for a 33\% subsampling
rate}
\end{figure}

\begin{figure}[tbh]
\includegraphics[width=8.5cm]{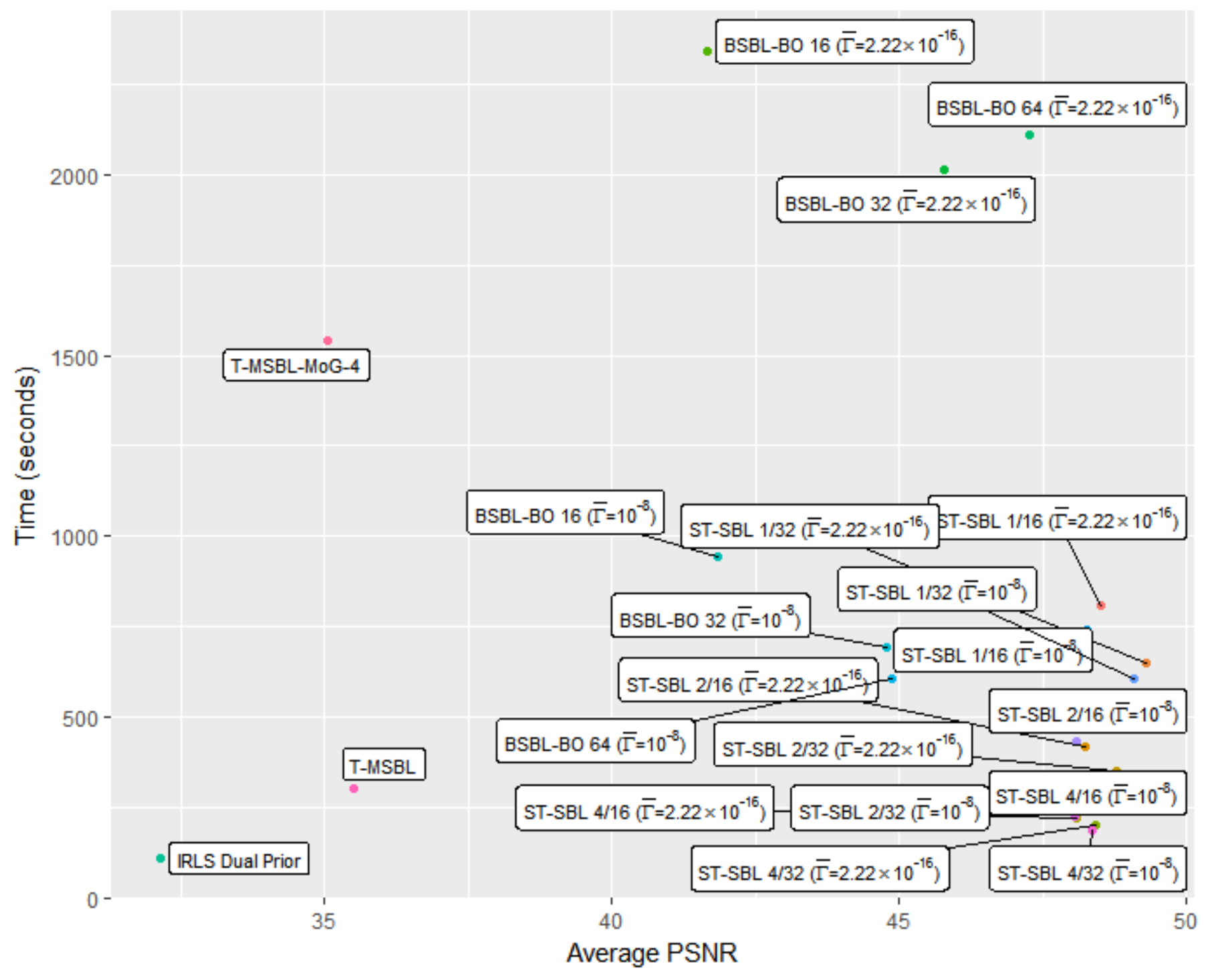}

\protect\caption{\label{fig:Time-vs-Performance50}Time vs Performance for a 50\% subsampling
rate}

\end{figure}

Figure \ref{fig:Image150results} shows Image 1 after being subsampled
at 50\% and then recovered with various algorithms. Figure \ref{fig:Image233results}
shows the same for Image 2 after it was subsampled at 33\%.

Whilst we are primarily concerned with the quality of the reconstruction,
the time taken for various reconstruction methods is of some practical
interest. Figures \ref{fig:Time-vs-Performance33} and \ref{fig:Time-vs-Performance50}
show plots of average time against the average PSNR achieved. For
these timings, the algorithms were run using MATLAB 2014a on computer
equipped with an i7-3770 processor and 8GB of RAM. Although processing
more than a single column at a time with ST-SBL does not typically
lead to improved recovery performance, the drop in performance is
minimal, and as seen in figures \ref{fig:Time-vs-Performance33} and
\ref{fig:Time-vs-Performance50}, there is a significant reduction
in computational time required by processing multiple columns at once.

Overall, the best recovery performance was achieved with ST-SBL 1/32
($\bar{\Gamma}=2.22\times10^{-16}$) . However, once we take the computational
time into account, we would consider ST-SBL 4/32 ($\bar{\Gamma}=10^{-8}$)
to be the best practical choice, as it offers only a minimal decrease
in performance, and a very significant reduction in the computational
time required. 

A number of other methods were also tested on this set of images.
The settings used for these methods are described below. 
\begin{itemize}
\item CoSaMP, FBMP and Subspace Pursuit were fed with the ``true'' number
of non-zero elements. This was chosen as the size of the support that
was used in the IRLS Dual Prior method. 
\item The value of $p$ for the calculation of the weights in the BIRLS
method was chosen in the same way as for IRLS.
\item BIRLS and Block OMP (BOMP) were tested with block sizes of 16 and
32.
\item MFOCUSS, MSBL, FOCUSS, FBMP, and $\ell$1-magic were fed with a small
value ($10^{-8}$) for the noise variance.
\end{itemize}
\begin{table}[H]
\begin{centering}
\begin{tabular}{|>{\centering}p{1.5cm}|>{\centering}p{0.6cm}|>{\centering}p{0.6cm}|>{\centering}p{0.6cm}|>{\centering}p{0.6cm}|>{\centering}p{0.6cm}|>{\centering}p{0.6cm}|>{\centering}p{0.6cm}|}
\hline 
\selectlanguage{british}%
{\footnotesize{}Image}\selectlanguage{english}%
 & {\footnotesize{}1} & {\footnotesize{}2} & {\footnotesize{}3} & {\footnotesize{}4} & {\footnotesize{}5} & {\footnotesize{}6} & {\footnotesize{}7}\tabularnewline
\hline 
\hline 
{\footnotesize{}HTP}{\footnotesize \par}

{\footnotesize{}\cite{foucart2011hard}} & {\footnotesize{}28.52} & {\footnotesize{}25.13} & {\footnotesize{}25.74} & {\footnotesize{}26.70} & {\footnotesize{}23.97} & {\footnotesize{}27.62} & {\footnotesize{}25.99}\tabularnewline
\hline 
\hline 
{\footnotesize{}EM-SBL}{\footnotesize \par}

{\footnotesize{}\cite{wipf2004sparse}} & {\footnotesize{}28.65} & {\footnotesize{}25.26} & {\footnotesize{}26.53} & {\footnotesize{}28.51} & {\footnotesize{}27.77} & {\footnotesize{}29.51} & {\footnotesize{}26.69}\tabularnewline
\hline 
\hline 
{\footnotesize{}FOCUSS\cite{gorodnitsky1997sparse}} & {\footnotesize{}27.75} & {\footnotesize{}24.62} & {\footnotesize{}25.98} & {\footnotesize{}27.43} & {\footnotesize{}25.38} & {\footnotesize{}27.65} & {\footnotesize{}26.43}\tabularnewline
\hline 
\hline 
{\footnotesize{}CoSaMP \cite{needell2009cosamp}} & {\footnotesize{}16.62} & {\footnotesize{}15.21} & {\footnotesize{}15.37} & {\footnotesize{}14.83} & {\footnotesize{}13.97} & {\footnotesize{}13.83} & {\footnotesize{}16.78}\tabularnewline
\hline 
\hline 
{\footnotesize{}Sl$0$ }{\footnotesize \par}

{\footnotesize{}\cite{mohimani2007fast}} & {\footnotesize{}30.57} & {\footnotesize{}26.70} & {\footnotesize{}27.89} & {\footnotesize{}29.46} & {\footnotesize{}29.33} & {\footnotesize{}31.88} & {\footnotesize{}27.78}\tabularnewline
\hline 
\hline 
{\footnotesize{}FBMP}{\footnotesize \par}

{\footnotesize{}\cite{schniter2008fast}} & {\footnotesize{}29.14} & {\footnotesize{}25.49} & {\footnotesize{}26.26} & {\footnotesize{}27.80} & {\footnotesize{}28.03} & {\footnotesize{}30.81} & {\footnotesize{}26.17}\tabularnewline
\hline 
\hline 
{\footnotesize{}Subspace Pursuit \cite{dai2008subspace}} & {\footnotesize{}28.14} & {\footnotesize{}24.92} & {\footnotesize{}24.73} & {\footnotesize{}24.85} & {\footnotesize{}22.68} & {\footnotesize{}28.52} & {\footnotesize{}25.09}\tabularnewline
\hline 
\hline 
{\footnotesize{}ExCoV}{\footnotesize \par}

{\footnotesize{}\cite{qiu2010variance}} & {\footnotesize{}25.17} & {\footnotesize{}22.72} & {\footnotesize{}23.80} & {\footnotesize{}22.64} & {\footnotesize{}22.91} & {\footnotesize{}22.51} & {\footnotesize{}24.35}\tabularnewline
\hline 
\hline 
{\footnotesize{}AMP }{\footnotesize \par}

{\footnotesize{}\cite{donoho2010message}} & {\footnotesize{}29.06} & {\footnotesize{}25.84} & {\footnotesize{}27.02} & {\footnotesize{}28.66} & {\footnotesize{}27.03} & {\footnotesize{}29.67} & {\footnotesize{}27.32}\tabularnewline
\hline 
\hline 
{\footnotesize{}BCS}{\footnotesize \par}

{\footnotesize{}\cite{ji2008bayesian}} & {\footnotesize{}28.36} & {\footnotesize{}24.91} & {\footnotesize{}25.71} & {\footnotesize{}27.26} & {\footnotesize{}27.40} & {\footnotesize{}30.20} & {\footnotesize{}25.72}\tabularnewline
\hline 
\hline 
{\footnotesize{}l1-magic }{\footnotesize \par}

{\footnotesize{}\cite{candes2005l1}} & {\footnotesize{}29.04} & {\footnotesize{}25.82} & {\footnotesize{}27.00} & {\footnotesize{}28.68} & {\footnotesize{}27.19} & {\footnotesize{}29.67} & {\footnotesize{}27.31}\tabularnewline
\hline 
\hline 
{\footnotesize{}BIRLS 16} & {\footnotesize{}38.17} & {\footnotesize{}34.97} & {\footnotesize{}35.60} & {\footnotesize{}36.64} & {\footnotesize{}38.25} & {\footnotesize{}40.18} & {\footnotesize{}36.14}\tabularnewline
\hline 
\hline 
\selectlanguage{british}%
{\footnotesize{}BOMP 16 \cite{eldar2010block}}\selectlanguage{english}%
 & \selectlanguage{british}%
{\footnotesize{}18.12}\selectlanguage{english}%
 & \selectlanguage{british}%
{\footnotesize{}16.88}\selectlanguage{english}%
 & \selectlanguage{british}%
{\footnotesize{}15.97}\selectlanguage{english}%
 & \selectlanguage{british}%
{\footnotesize{}16.56}\selectlanguage{english}%
 & \selectlanguage{british}%
{\footnotesize{}14.26}\selectlanguage{english}%
 & \selectlanguage{british}%
{\footnotesize{}13.45}\selectlanguage{english}%
 & \selectlanguage{british}%
{\footnotesize{}18.20}\selectlanguage{english}%
\tabularnewline
\hline 
\hline 
{\footnotesize{}BIRLS 32} & \selectlanguage{british}%
\textbf{\footnotesize{}39.28}\selectlanguage{english}%
 & \selectlanguage{british}%
\textbf{\footnotesize{}36.68}\selectlanguage{english}%
 & \selectlanguage{british}%
\textbf{\footnotesize{}36.68}\selectlanguage{english}%
 & \selectlanguage{british}%
\textbf{\footnotesize{}37.79}\selectlanguage{english}%
 & \selectlanguage{british}%
\textbf{\footnotesize{}38.95}\selectlanguage{english}%
 & \selectlanguage{british}%
\textbf{\footnotesize{}41.09}\selectlanguage{english}%
 & \selectlanguage{british}%
\textbf{\footnotesize{}37.44}\selectlanguage{english}%
\tabularnewline
\hline 
\hline 
\selectlanguage{british}%
{\footnotesize{}BOMP 32}\selectlanguage{english}%
 & \selectlanguage{british}%
{\footnotesize{}17.76}\selectlanguage{english}%
 & \selectlanguage{british}%
{\footnotesize{}16.24}\selectlanguage{english}%
 & \selectlanguage{british}%
{\footnotesize{}16.48}\selectlanguage{english}%
 & \selectlanguage{british}%
{\footnotesize{}15.81}\selectlanguage{english}%
 & \selectlanguage{british}%
{\footnotesize{}14.44}\selectlanguage{english}%
 & \selectlanguage{british}%
{\footnotesize{}14.49}\selectlanguage{english}%
 & \selectlanguage{british}%
{\footnotesize{}17.47}\selectlanguage{english}%
\tabularnewline
\hline 
\hline 
\selectlanguage{british}%
{\footnotesize{}MSBL }{\footnotesize \par}

{\footnotesize{}\cite{wipf2007empirical}}\selectlanguage{english}%
 & \selectlanguage{british}%
{\footnotesize{}33.12}\selectlanguage{english}%
 & \selectlanguage{british}%
{\footnotesize{}29.13}\selectlanguage{english}%
 & \selectlanguage{british}%
{\footnotesize{}29.67}\selectlanguage{english}%
 & \selectlanguage{british}%
{\footnotesize{}30.66}\selectlanguage{english}%
 & \selectlanguage{british}%
{\footnotesize{}32.77}\selectlanguage{english}%
 & \selectlanguage{british}%
{\footnotesize{}34.65}\selectlanguage{english}%
 & \selectlanguage{british}%
{\footnotesize{}30.01}\selectlanguage{english}%
\tabularnewline
\hline 
\hline 
\selectlanguage{british}%
{\footnotesize{}MFOCUSS\cite{cotter2005sparse}}\selectlanguage{english}%
 & \selectlanguage{british}%
{\footnotesize{}33.30}\selectlanguage{english}%
 & \selectlanguage{british}%
{\footnotesize{}29.99}\selectlanguage{english}%
 & \selectlanguage{british}%
{\footnotesize{}31.04}\selectlanguage{english}%
 & \selectlanguage{british}%
{\footnotesize{}32.30}\selectlanguage{english}%
 & \selectlanguage{british}%
{\footnotesize{}34.11}\selectlanguage{english}%
 & \selectlanguage{british}%
{\footnotesize{}35.77}\selectlanguage{english}%
 & \selectlanguage{british}%
{\footnotesize{}30.57}\selectlanguage{english}%
\tabularnewline
\hline 
\end{tabular}
\par\end{centering}

\protect\caption{\label{tab:set2multi33}Reconstruction quality of image set 2 after
the DCT was subsampled at the 50\% level ($\mathbf{A}\in\mathbb{R}^{171\times512}$).
Results given in terms of PSNR.}
\end{table}

\begin{table}[H]
\begin{centering}
\begin{tabular}{|>{\centering}p{1.5cm}|>{\centering}p{0.6cm}|>{\centering}p{0.6cm}|>{\centering}p{0.6cm}|>{\centering}p{0.6cm}|>{\centering}p{0.6cm}|>{\centering}p{0.6cm}|>{\centering}p{0.6cm}|}
\hline 
 & {\footnotesize{}1} & {\footnotesize{}2} & {\footnotesize{}3} & {\footnotesize{}4} & {\footnotesize{}5} & {\footnotesize{}6} & {\footnotesize{}7}\tabularnewline
\hline 
{\footnotesize{}HTP} & {\footnotesize{}21.63} & {\footnotesize{}19.17} & {\footnotesize{}19.29} & {\footnotesize{}19.91} & {\footnotesize{}18.83} & {\footnotesize{}20.21} & {\footnotesize{}20.46}\tabularnewline
\hline 
{\footnotesize{}EM-SBL} & {\footnotesize{}22.67} & {\footnotesize{}FAIL} & {\footnotesize{}FAIL} & {\footnotesize{}FAIL} & {\footnotesize{}FAIL} & {\footnotesize{}FAIL} & {\footnotesize{}FAIL}\tabularnewline
\hline 
{\footnotesize{}FOCUSS} & {\footnotesize{}22.20} & {\footnotesize{}20.00} & {\footnotesize{}20.87} & {\footnotesize{}21.07} & {\footnotesize{}20.07} & {\footnotesize{}20.69} & {\footnotesize{}21.91}\tabularnewline
\hline 
{\footnotesize{}CoSaMP} & {\footnotesize{}17.00} & {\footnotesize{}15.54} & {\footnotesize{}15.81} & {\footnotesize{}15.48} & {\footnotesize{}13.75} & {\footnotesize{}14.69} & {\footnotesize{}17.54}\tabularnewline
\hline 
{\footnotesize{}Sl$0$} & {\footnotesize{}23.17} & {\footnotesize{}FAIL} & {\footnotesize{}FAIL} & {\footnotesize{}FAIL} & {\footnotesize{}FAIL} & {\footnotesize{}FAIL} & {\footnotesize{}FAIL}\tabularnewline
\hline 
{\footnotesize{}FBMP} & {\footnotesize{}23.51} & {\footnotesize{}20.88} & {\footnotesize{}21.96} & {\footnotesize{}24.04} & {\footnotesize{}20.99} & {\footnotesize{}24.56} & {\footnotesize{}22.10}\tabularnewline
\hline 
{\footnotesize{}Subspace Pursuit} & {\footnotesize{}20.78} & {\footnotesize{}12.73} & {\footnotesize{}12.89} & {\footnotesize{}9.551} & {\footnotesize{}9.123} & {\footnotesize{}21.04} & {\footnotesize{}14.36}\tabularnewline
\hline 
{\footnotesize{}ExCoV} & {\footnotesize{}24.36} & {\footnotesize{}FAIL} & {\footnotesize{}FAIL} & {\footnotesize{}FAIL} & {\footnotesize{}FAIL} & {\footnotesize{}FAIL} & {\footnotesize{}FAIL}\tabularnewline
\hline 
{\footnotesize{}AMP} & {\footnotesize{}22.97} & {\footnotesize{}20.71} & {\footnotesize{}21.67} & {\footnotesize{}22.10} & {\footnotesize{}20.78} & {\footnotesize{}21.85} & {\footnotesize{}22.48}\tabularnewline
\hline 
{\footnotesize{}BCS} & {\footnotesize{}22.59} & {\footnotesize{}20.22} & {\footnotesize{}20.66} & {\footnotesize{}22.73} & {\footnotesize{}20.06} & {\footnotesize{}22.89} & {\footnotesize{}21.06}\tabularnewline
\hline 
{\footnotesize{}l1-magic} & {\footnotesize{}22.93} & {\footnotesize{}20.65} & {\footnotesize{}21.64} & {\footnotesize{}22.09} & {\footnotesize{}20.75} & {\footnotesize{}21.85} & {\footnotesize{}22.48}\tabularnewline
\hline 
{\footnotesize{}BIRLS 16} & {\footnotesize{}31.71} & {\footnotesize{}27.92} & {\footnotesize{}28.75} & {\footnotesize{}30.62} & {\footnotesize{}30.35} & {\footnotesize{}31.68} & {\footnotesize{}30.10}\tabularnewline
\hline 
\selectlanguage{british}%
{\footnotesize{}BOMP 16}\selectlanguage{english}%
 & \selectlanguage{british}%
{\footnotesize{}17.42}\selectlanguage{english}%
 & \selectlanguage{british}%
{\footnotesize{}15.52}\selectlanguage{english}%
 & \selectlanguage{british}%
{\footnotesize{}15.71}\selectlanguage{english}%
 & \selectlanguage{british}%
{\footnotesize{}16.95}\selectlanguage{english}%
 & \selectlanguage{british}%
{\footnotesize{}16.68}\selectlanguage{english}%
 & \selectlanguage{british}%
{\footnotesize{}16.88}\selectlanguage{english}%
 & \selectlanguage{british}%
{\footnotesize{}18.24}\selectlanguage{english}%
\tabularnewline
\hline 
{\footnotesize{}BIRLS 32} & \selectlanguage{british}%
\textbf{\footnotesize{}32.18}\selectlanguage{english}%
 & \selectlanguage{british}%
\textbf{\footnotesize{}28.34}\selectlanguage{english}%
 & \selectlanguage{british}%
\textbf{\footnotesize{}29.19}\selectlanguage{english}%
 & \selectlanguage{british}%
\textbf{\footnotesize{}31.27}\selectlanguage{english}%
 & \selectlanguage{british}%
\textbf{\footnotesize{}30.97}\selectlanguage{english}%
 & \selectlanguage{british}%
\textbf{\footnotesize{}32.17}\selectlanguage{english}%
 & \selectlanguage{british}%
\textbf{\footnotesize{}30.43}\selectlanguage{english}%
\tabularnewline
\hline 
\selectlanguage{british}%
{\footnotesize{}BOMP 32}\selectlanguage{english}%
 & \selectlanguage{british}%
{\footnotesize{}14.59}\selectlanguage{english}%
 & \selectlanguage{british}%
{\footnotesize{}13.80}\selectlanguage{english}%
 & \selectlanguage{british}%
{\footnotesize{}13.12}\selectlanguage{english}%
 & \selectlanguage{british}%
{\footnotesize{}13.25}\selectlanguage{english}%
 & \selectlanguage{british}%
{\footnotesize{}14.48}\selectlanguage{english}%
 & \selectlanguage{british}%
{\footnotesize{}14.09}\selectlanguage{english}%
 & \selectlanguage{british}%
{\footnotesize{}14.60}\selectlanguage{english}%
\tabularnewline
\hline 
\selectlanguage{british}%
{\footnotesize{}MSBL}\selectlanguage{english}%
 & \selectlanguage{british}%
{\footnotesize{}29.82}\selectlanguage{english}%
 & \selectlanguage{british}%
{\footnotesize{}26.03}\selectlanguage{english}%
 & \selectlanguage{british}%
{\footnotesize{}27.21}\selectlanguage{english}%
 & \selectlanguage{british}%
{\footnotesize{}28.51}\selectlanguage{english}%
 & \selectlanguage{british}%
{\footnotesize{}28.90}\selectlanguage{english}%
 & \selectlanguage{british}%
{\footnotesize{}31.03}\selectlanguage{english}%
 & \selectlanguage{british}%
{\footnotesize{}28.28}\selectlanguage{english}%
\tabularnewline
\hline 
\selectlanguage{british}%
{\footnotesize{}MFOCUSS}\selectlanguage{english}%
 & \selectlanguage{british}%
{\footnotesize{}30.98}\selectlanguage{english}%
 & \selectlanguage{british}%
{\footnotesize{}27.15}\selectlanguage{english}%
 & \selectlanguage{british}%
{\footnotesize{}28.87}\selectlanguage{english}%
 & \selectlanguage{british}%
{\footnotesize{}29.86}\selectlanguage{english}%
 & \selectlanguage{british}%
{\footnotesize{}30.37}\selectlanguage{english}%
 & \selectlanguage{british}%
{\footnotesize{}32.07}\selectlanguage{english}%
 & \selectlanguage{british}%
{\footnotesize{}28.66}\selectlanguage{english}%
\tabularnewline
\hline 
\end{tabular}
\par\end{centering}

\protect\caption{\label{tab:set2multi50Reconstruction-quality-of}Reconstruction quality
of image set 2 after the DCT was subsampled at the 33\% level ($\mathbf{A}\in\mathbb{R}^{256\times512}$).
Results given in terms of PSNR. FAIL indicates that the method returned
only zeros.}
\end{table}

\selectlanguage{british}%
The results of these tests can be seen in Table \ref{tab:set2multi50Reconstruction-quality-of},
which shows the results when the subsampling rate was 50\%, and Table
\ref{tab:set2multi33} which shows the results when the subsampling
rate was 33\%. The only method that performed sufficiently well to
be of interest was BIRLS, which was the only method to outperform
any of the previously tested methods. However, it was still outperformed
by the previously tested methods that took advantage of block structure.
This suggests that the use of the block structure assumption is able
to provide significant performance gains. It is also notable that
at the 33\% subsampling level, some of the methods returned a vector
consisting only of zeros when attempting to recover some images, and
in these cases the methods are considered to have failed entirely.

However, block sparsity is clearly not a sufficient assumption in
and of itself, which can be seen clearly from the dreadful performance
of the \foreignlanguage{english}{BOMP} method. The easiest explanation
for this is that due to the nature of the method, a number of elements
of the estimated vector are guaranteed to be zero, and the effect
this has on performance can also be seen from the poor performance
of the CoSaMP method. 

Also of interest is the fact that both \foreignlanguage{english}{MFOCUSS}
and \foreignlanguage{english}{MSBL} outperformed their \foreignlanguage{english}{SMV
}model counterparts. Unlike ST-SBL, these methods do not make any
use of block structure. This suggests that attempting to use both
block structure and joint sparsity on the signal effectively forces
``too much'' structure on the signal, leading to worse performance.
This is in line with the poor results of \foreignlanguage{english}{BOMP},
which forces some blocks to be exactly zero, whereas ST-SBL, \foreignlanguage{english}{BSBL-BO,}
and \foreignlanguage{english}{BIRLS} all allow for solutions to be
only approximately sparse.

\subsection{Brno dataset}

Due to the large size of this test set, only a limited number of algorithms
were tested. The algorithms that were chosen for testing are ST-SBL\foreignlanguage{english}{
1/32 ($\bar{\Gamma}=2.22\times10^{-16}$)}, ST-SBL \foreignlanguage{english}{4/32
($\bar{\Gamma}=10^{-8}$)}, \foreignlanguage{english}{BIRLS} and $\ell$1-magic

The results in terms of average PSNRand the number of times the method
returned the best PSNR can be seen in Table \ref{tab:33set3} for
downsampling at the 33\% level, and Table \ref{tab:50set3} for downsampling
at the 50\% level.

\begin{table}[H]
\begin{centering}
\begin{tabular}{|>{\centering}p{2.5cm}|>{\centering}p{2cm}|>{\centering}p{3cm}|}
\hline 
\selectlanguage{english}%
\selectlanguage{british}%
 & Average PSNR  & Number of times best method\tabularnewline
\hline 
\hline 
\selectlanguage{english}%
ST-SBL 1/32 ($\bar{\Gamma}=2.22\times10^{-16}$)\selectlanguage{british}%
 & 24.90 & 20\tabularnewline
\hline 
\selectlanguage{english}%
ST-SBL 4/32 ($\bar{\Gamma}=10^{-8}$)\selectlanguage{british}%
 & \textbf{25.25} & \textbf{59}\tabularnewline
\hline 
BIRLS 32 & 23.46 & 5\tabularnewline
\hline 
$\ell$1-magic & 12.47 & 0\tabularnewline
\hline 
\end{tabular}
\par\end{centering}

\protect\caption{\label{tab:33set3}Reconstruction quality of images from set 3 after
downsampling at the 33\% level}
\end{table}

\begin{table}[H]
\begin{centering}
\begin{tabular}{|>{\centering}p{2.5cm}|>{\centering}p{2cm}|>{\centering}p{3cm}|}
\hline 
\selectlanguage{english}%
\selectlanguage{british}%
 & Average PSNR  & Number of times best method\tabularnewline
\hline 
\hline 
\selectlanguage{english}%
ST-SBL 1/32 ($\bar{\Gamma}=2.22\times10^{-16}$)\selectlanguage{british}%
 & 24.96 & 20\tabularnewline
\hline 
\selectlanguage{english}%
ST-SBL 4/32 ($\bar{\Gamma}=10^{-8}$)\selectlanguage{british}%
 & \textbf{25.31} & \textbf{62}\tabularnewline
\hline 
BIRLS 32 & 23.44 & 2\tabularnewline
\hline 
$\ell$1-magic & 16.53 & 0\tabularnewline
\hline 
\end{tabular}
\par\end{centering}

\protect\caption{\label{tab:50set3}Reconstruction quality of images from set 3 after
downsampling at the 50\% level}
\end{table}

The results are mostly as would be expected from previous results.
The only unexpected item to note is that \foreignlanguage{english}{ST-SBL}
\foreignlanguage{english}{4/32 ($\bar{\Gamma}=10^{-8}$)} is outperforming
\foreignlanguage{english}{ST-SBL} \foreignlanguage{english}{1/32 ($\bar{\Gamma}=2.22\times10^{-16}$))}
on this data set. This could be due to the signals being downsampled
after envelope detection and logarithmic correction, or it could be
due to the support of the DCTs of neighbouring ultrasound lines being
more closely related in this set of images.

\selectlanguage{english}%

\section{Conclusions}

\label{sec:conclusions}

In this paper, we have thoroughly investigated what we consider to
be the state-of-the-art methods for the reconstruction of compressively
sensed medical ultrasound images. We have shown that by varying the
parameters of structured Sparse Bayesian learning methods, we can
achieve significant improvements in the recovery of compressively
sensed ultrasound images. On the other hand, if we are willing to
accept a slight decrease in recovery performance, we can significantly
reduce the computational time required for recovery. The advantage
of the structured Sparse Bayesian Learning methods is very significant,
and it is worth considering why this is the case. The IRLS approach
is an attempt to improve upon $\ell_{1}$ norm minimisation by more
closely mimicking $\ell_{0}$ pseudonorm minimisation. However, if
we examine the signals we are sensing, we see that in fact they have
very few non-zero elements, and hence $\ell_{0}$ pseudonorm minimisation
is unlikely to be the ideal approach. Therefore, we can conclude that
this advantage comes from the previously noted ability of these methods
to recover non-sparse signals \cite{zhang2014spatiotemporal}. Hence
although it may be of some theoretical interest, development of methods
which more closely approximate $\ell_{0}$ pseudonorm minimisation
is unlikely to provide significant practical advantages, and efforts
to improve the reconstruction of compressively sensed signals should
instead focus on more accurately modeling the structure and statistical
properties of the signals.

Our current work focuses on taking advantage of the statistical properties
of ultrasound images in the MMV case, building on previous work \cite{achim2010compressive,achim2014reconstruction}.

\section{Acknowledgements}

The authors would like to thank Dr Zhilin Zhang from Samsung Research
America - Dallas for making his code for T-MSBL and BSBL available
online, and for providing us with the code for the ST-SBL method.
We would also like to thank Dr Adrian Basarab from IRIT Laboratory,
Toulouse, France for providing the thyroid ultrasound images used
in this study.

\noindent This work was supported in part by the Engineering and Physical
Sciences Research Council (EP/I028153/1) and the University of Bristol.

\bibliographystyle{ieeetr}
\bibliography{Bibliography}

\end{document}